\UseRawInputEncoding
\documentclass[twocolumn,superscriptaddress,amssymb,amsmath,aps,prl]{revtex4-2}

\usepackage{appendix}
\usepackage{amsmath}
\usepackage{graphicx}% Include figure files
\usepackage{color}

\usepackage{soul}
\usepackage{dcolumn}% Align table columns on decimal point
\usepackage{bm}% bold math
\usepackage{hyperref}
\hypersetup{
    colorlinks=true,
    linkcolor=blue,
    filecolor=magenta,      
    urlcolor=cyan,
}
\usepackage{xr}
\usepackage{subfigure}

\newcommand{\LJ}[1]{\textcolor{red}{#1}}

\begin{document}

%\preprint{APS/123-QED}

\title{Engineering fast bias-preserving gates on stabilized cat qubits}% Force line breaks with \\

\author{Qian Xu}
\affiliation{Pritzker School of Molecular Engineering, The University of Chicago, Chicago 60637, USA}

\author{Joseph K Iverson}
%\affiliation{IQIM, California Institude of Technology, Pasadena, CA 91125, USA}
\affiliation{AWS Center for Quantum Computing, Pasadena, CA 91125, USA}

\author{Fernando G.S.L. Brand\~ao}
\affiliation{AWS Center for Quantum Computing, Pasadena, CA 91125, USA}
\affiliation{IQIM, California Institude of Technology, Pasadena, CA 91125, USA}

\author{Liang Jiang}
\email{liang.jiang@uchicago.edu}
\affiliation{Pritzker School of Molecular Engineering, The University of Chicago, Chicago 60637, USA}
\affiliation{AWS Center for Quantum Computing, Pasadena, CA 91125, USA}

\date{\today}% It is always \today, today,

\begin{abstract}
% It has been recently proposed that cat qubits, which possesses a biased noise channel, can be stabilized in a driven Kerr oscillator. A set of gates on the cat qubits, including a controlled-NOT gate, can be constructed in a way that preserves the noise bias. In the presence of photo loss, the gates have to be implemented fast enough in order to obtain high gate fidelity. However, as the gate speed increases the non-adiabaticity of the gates causes leakage from the codespace, which after being projected back induces the minor type of error and destroys the noise bias. We show that adding additional engineered two-photon dissipation helps suppress the minor type of error but enhances the major type of error at the same time. To address this problem, we apply shortcuts to adiabaticity (STA) methods to the originally proposed gates without two-photon dissipation to suppress the non-adiabatic errors so that additional two-photon dissipation during the implementation of the gates is no longer in need. We show that using the improved control scheme we can obtain higher gate fidelity and higher noise bias simultaneously in the presence of a realistic level of noise, which can significantly reduce the resource overhead required for concatenating the cat qubits with a second level of coding to implement concatenated error correction.

Stabilized cat codes can provide a biased noise channel with a set of bias-preserving (BP) gates, which can significantly reduce the resource overhead for fault-tolerant quantum computing. 
All existing schemes of BP gates, however, require adiabatic quantum evolution, with performance limited by excitation loss and non-adiabatic errors during the adiabatic gates. 
%A two-component cat qubit can be stabilized in a driven Kerr nonlinear oscillator or by engineered driven two-photon dissipation. Such stabilized cat qubits possess a biased noise channel and a set of bias-preserving (BP) gates have been separately proposed on these two type of cats. We observe that although both limited by the non-adiabatic effects, compared to the dissipative cat, the Kerr cat supports faster gate operations with higher gate fidelity due to the unitary nature of the gates. However, the existing control for these unitary gates may not be able to achieve noise bias as good as the dissipative gates do. 
In this work, we apply a derivative-based leakage suppression technique to overcome non-adiabatic errors, so that we can implement fast BP gates on Kerr-cat qubits with improved gate fidelity while maintaining high noise bias. 
%We show that the improved gates, when applied to concatenated quantum error correction, can lead to lower logical error rate. 
%Based on our result, we propose to use an architecture that hybrids the Kerr nonlinearity with two-photon dissipation to better protect the qubit. In this architecture, the two-photon dissipation should be turned on when the cat is idling to keep the system sufficiently cool while turned off when implementing the the BP gates to enable high-fidelity operations.
When applied to concatenated quantum error correction, the fast BP gates can not only improve the logical error rate but also reduce resource overhead, which enables more efficient implementation of fault-tolerant quantum computing.
\end{abstract}

\pacs{Valid PACS appear here}% PACS, the Physics and Astronomy
\maketitle

%\tableofcontents

% \section{\label{sec:level1}Introduction}
%Quantum error correction (QEC) provides a solution to protect quantum information in the presence of decoherence and various imperfections. 
Quantum error correction (QEC) of generic errors is very challenging, because of the demanding threshold requirements and significant resource overhead. 
To overcome this challenge, we may adaptively design the QEC codes targeting practically relevant errors, so that we can correct more errors in a hardware-efficient way. 
For example, we can develop various efficient bosonic QEC codes to protect encoded quantum information from excitation loss errors \cite{GKP, Leghtas13b, Michael16, Albert18a},
which have been experimentally demonstrated using superconducting circuits \cite{Ofek16, HuL19, Lescanne20, Grimm20, Campagne20} and trapped ions \cite{Fluhmann19}.

With Hamiltonian protection or reservoir engineering, some bosonic codes can continuously suppress practically relevant errors (e.g., dephasing, excitation loss, and etc) and also provide a highly biased noise channel. For example, stabilized cat qubits can exponentially suppress bit-flip errors at the encoded level, because of the large separation of coherent states in the phase space \cite{Mirrahimi14, Lescanne20, Grimm20}. 
Such an encoding with highly biased noise channel can play a unique role in fault-tolerant architecture \cite{tuckett2018ultrahigh,bonilla2020xzzx}, as the higher-level QEC codes can be tailored toward the biased noise to exhibit significantly improved error threshold and resource overhead. 
To get the maximum benefit from biased noise, it is essential for all gate operations to consistently preserve the noise bias. 
Recently, a non-trivial set of bias-preserving (BP) gates have been discovered for stabilized cat qubits \cite{puri2017engineering,Puri20, Grimm20, Guillaud_Mirrahimi_2019}, which opens up a new direction of fault-tolerant architectural design \cite{Guillaud_Mirrahimi_2020, Chamberland_2020, darmawan2021practical}.
All existing schemes of BP gates, however, require adiabatic quantum evolution, with performance limited by excitation loss and non-adiabatic errors during the slow adiabatic gates. Therefore, it is desirable to develop fast BP gates to suppress non-adiabatic errors while preserving high noise bias.

In this work, we propose fast BP gates on stabilized cat qubits with unitary control to coherently suppress non-adiabatic errors. We systematically compare the performance of fast BP gates with existing adiabatic BP gates \cite{Guillaud_Mirrahimi_2019, Puri20}, in terms of the gate fidelity, noise bias, and concatenated logical gate fidelity.

 \begin{figure}[t]
    \centering
    \includegraphics[width = 0.5\textwidth]{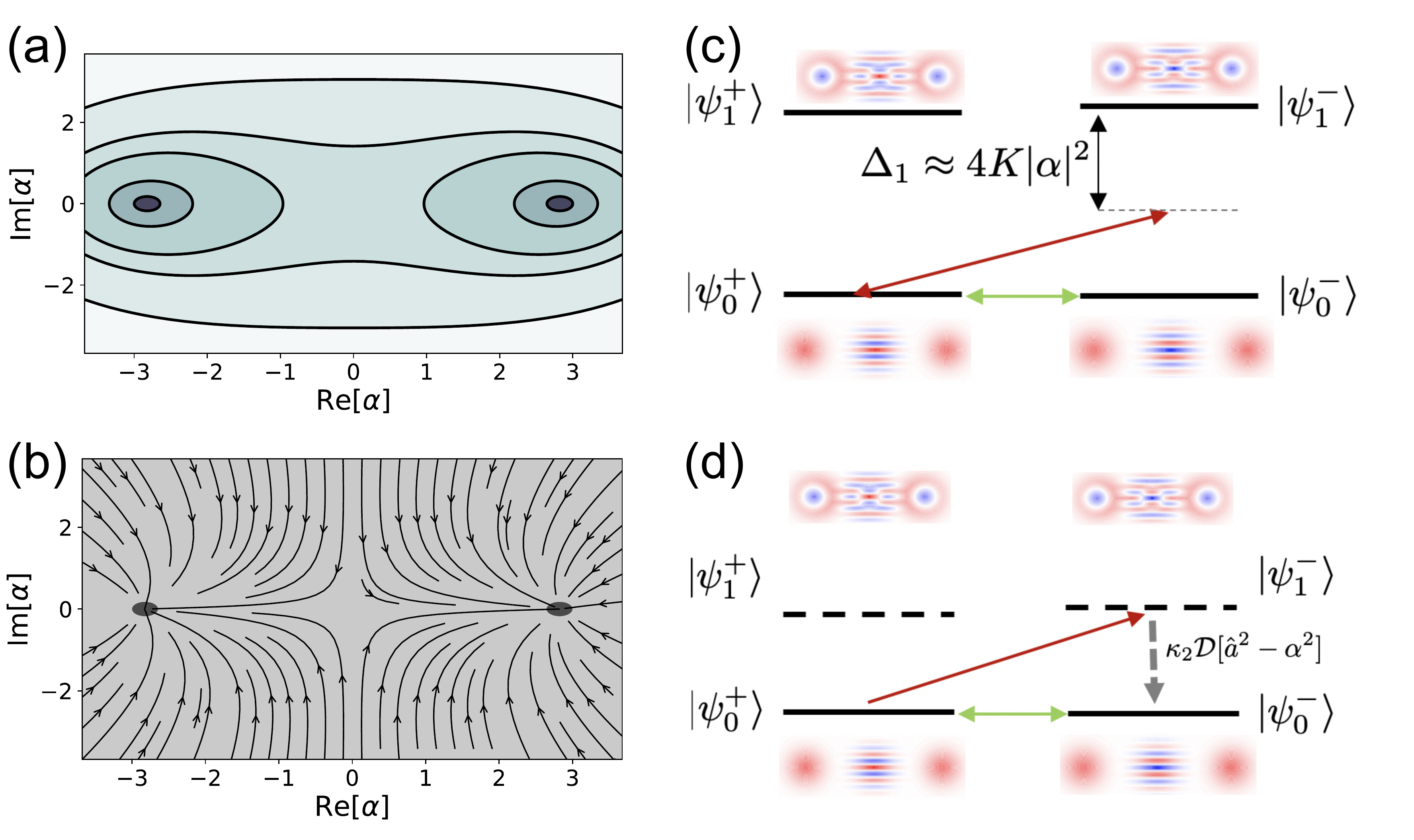}
    % \caption{(a). The semi-classical potential $\langle \hat{H}_{KPO}\rangle$ of a Kerr cat in phase space. (b). The semi-classical dynamics of a dissipative cat in phase space. (c)(d). Mechanism of the non-adiabatic errors during the Z-axis rotation on (c) Kerr cats and (d) dissipative cats. $|\psi_0^{\pm}\rangle, |\psi_1^{\pm}\rangle$ denote the first two pairs of eigenstates of $\hat{H}_{KPO}$ with parity $\pm$ and the spacing between states in the vertical axis in (c) and (d) represents the energy gap and dissipation gap of the Kerr cat and the dissipative cat, respectively. For Kerr cats in (c), a linear drive ($\Omega(t)\hat{a} + h.c.$) not only couples the ground states $|\psi_0^{\pm}\rangle$ of $\hat{H}_{KPO}$ but also induces leakage outside the ground state manifold. Since the leakage transition is off-resonant by an energy gap $\Delta_1$ the non-adiabatic errors manifest in the form of off-resonant leakage. While for dissipative cats in (d) with a dissipation gap, the leakage-transition induces decoherence within the cat state manifold as the cat continuously loses pairs of photons to the engineered environment.}
     \caption{(a). The semi-classical potential $\langle \hat{H}_{KPO}\rangle$ of a Kerr cat in phase space. (b). The semi-classical dynamics of a dissipative cat in phase space. (c)(d). Mechanism of the non-adiabatic errors during the Z-axis rotation on (c) Kerr cats and (d) dissipative cats. $|\psi_0^{\pm}\rangle, |\psi_1^{\pm}\rangle$ denote the first two pairs of eigenstates of $\hat{H}_{KPO}$ with parity $\pm$ and the spacing between states in the vertical axis in (c) and (d) represents the energy gap and dissipative gap of the Kerr cat and the dissipative cat, respectively. For Kerr cats in (c), the non-adiabatic errors induced by the linear drive ($\Omega(t)\hat{a} + h.c.$) manifest in the form of off-resonant leakage. While for dissipative cats in (d) with a dissipation gap, the linear drive induces decoherence within the cat state manifold when combined with the engineered two-photon dissipation.}
    \label{fig:double_well_potential}
\end{figure}

%There are two approaches to stabilize the cat qubits, with a two dimensional logical space spanned by $|\alpha\rangle$ and $|-\alpha\rangle$. 
%%%%%%%%%%%%% 
The cat qubit spanned by coherent states $|\alpha\rangle$ and $|-\alpha\rangle$ can be stabilized in a Kerr oscillator with parametric two-photon drive \cite{puri2017engineering, Grimm20, Puri20, goto2016universal}. The Hamiltonian of such a Kerr parametric oscillator (KPO) in the frame rotating at the oscillator frequency is:
\begin{equation}
\begin{aligned}
\hat{H}_{KPO} = - K(\hat{a}^{2\dagger} - \alpha^2)(\hat{a}^2 - \alpha^2),\\
\end{aligned}
\label{eq:Kerr_Hamiltonian}
\end{equation}
where $K$ is the strength of Kerr nonlinearity. In the rotating frame, we may intuitively view the KPO system as a ``double-well" potential with two extrema $\alpha$ and $-\alpha$ in phase space, as shown in Fig.~\ref{fig:double_well_potential}(a). 
The ground state manifold has a two-fold degeneracy -- spanned by the Schrodinger Cat states $|\psi_0^{\pm}\rangle = \mathcal{N}_{\pm} (|\alpha\rangle \pm |-\alpha\rangle)$ with even and odd number of excitations, respectively. Similarly, the excited states $|\psi_n^{\pm}\rangle$ are nearly degenerate pairs with eigenenergies $\Delta_n \pm \delta_n/2$ for $n=1,2,\cdots$, where $\pm$ again labels the parity and $\delta_n$ denotes the energy splitting between the n-th pair, which is exponentially suppressed for large $\alpha^2$ provided that $\Delta_n$ is well below the potential barrier. The excitation gap $\Delta = \Delta_1 \approx - 4 K |\alpha|^2$ \footnote{Despite the energy gap $\Delta_1 <0$ is negative in the rotating frame, the excitation energy $\omega_0 + \Delta_1 >0$ is still positive in the lab frame, with oscillator frequency $\omega_0$.} provides continuous protection of the encoded quantum information. 

% Similarly, the Hamiltonian supports pairs of nearly degenerate excited eigenstates $|\psi_n^{\pm}\rangle$ with eigenenergies $\Delta_n \pm \delta_n/2$ for $n=1,2,\cdots$, where $\pm$ indicates the states with even and odd photon number parity respectively and $\delta_n$ denotes the energy splitting, which is exponentially suppressed for large $\alpha^2$ provided that $\Delta_n$ is well below the potential barrier. The excited states are gapped from the ground states by $\Delta \approx - 4 K |\alpha|^2$
% \footnote{Despite the energy gap $\Delta_1 <0$ is negative in the rotating frame, the excitation energy $\omega_0 + \Delta_1 >0$ is still positive in the lab frame, with oscillator frequency $\omega_0$.}, 
% which provides continuous protection of the encoded quantum information. 
%The cat subspace is stabilized by this large energy gap if there is no resonant excitation.

Alternatively, the cat qubit can also be stabilized by engineered two-photon dissipation and two-photon drive  \cite{Mirrahimi14, Guillaud_Mirrahimi_2019, Guillaud_Mirrahimi_2020}: 
\begin{equation}
\frac{d \rho}{d t}=\kappa_{2 } \mathcal{D}\left[\hat{a}^{2}-\alpha^{2}\right] \rho,
\label{eq:two-photon_dissipation}
\end{equation}
where $\kappa_2$ is the two-photon dissipation rate and 
$\mathcal{D}[\hat{A}]\hat{\rho} = \hat{A} \hat{\rho} \hat{A}^{\dagger} - \frac{1}{2} \{\hat{A}^{\dagger}\hat{A}, \hat{\rho}\}$. 
We may intuitively understand the stabilization using a semi-classical flow diagram with two stable steady states $|\pm\alpha\rangle$ as illustrated in Fig.~\ref{fig:double_well_potential}(b). For quantum evolution, the entire code space spanned by $|\pm\alpha\rangle$ is stabilized as an attractive steady-state subspace, which is protected by a dissipative gap.

% One can in principle hybrid these two type of stabilization to protect the cat qubit, i.e. applying two-photon dissipation to a strong Kerr nonlinear oscillator. And in fact, in the momery level, additional two-photon dissipation can be helpful to further cool the Kerr oscillator if, for example, the cat is coupled to a thermal bath with wide-band spectral density.\cite{Puri_St-Jean_Gross_Grimm_Frattini_Iyer_Krishna_Touzard_Jiang_Blais_et_al_2020}

We define the computational basis of the cat qubit as $|0\rangle_L \equiv \frac{|\psi_0^+\rangle + |\psi_0^-\rangle}{\sqrt{2}} \approx |\alpha\rangle, |1\rangle_L \equiv \frac{|\psi_0^+\rangle - |\psi_0^-\rangle}{\sqrt{2}} \approx |-\alpha\rangle$, where the inaccuracy of the approximation decreases exponentially with $\alpha^2$. Excitation loss error may change the number parity and induce phase flip error $\hat{Z}$, but the back action to the population of different coherent states decreases exponentially with $|\alpha|^2$. Hence, the noise channel of the stabilized cat qubit is strongly biased toward phase flip error $\hat{Z}$, with the noise bias $\eta \equiv P_z/P_x$ increasing exponentially with $|\alpha|^2$.
%provided that the leakage outside the cat logical space is sufficiently small. 
Here, and in the rest of the paper, we use $P_x$ to denote all the non-dephasing type of errors for simplicity.

A set of bias-preserving gates have been proposed separately for Kerr cat\cite{Puri20} and dissipative cat\cite{Guillaud_Mirrahimi_2019}, which we refer to as Kerr gates and dissipative gates, respectively. 
% All these schemes require the adiabatic implementation of the Z rotation ($\exp (i \theta \hat{Z} / 2)$), ZZ rotation ($\exp \left(i \theta \widehat{Z}_{1} \hat{Z}_{2} / 2\right)$) and CX (controlled-X) gate, with gate time  $T \gg 1 /|\Delta(\Delta_d)|$ where $\Delta(\Delta_d)$ is the energy (dissipation) gap protecting the cat qubit.
Under excitation loss, the total Z and X error probabilities of the BP gates can be written as the following:
\begin{equation}
\begin{array}{c}
P_z = P_z^{NA} + \beta \kappa_1 |\alpha|^2 T \\
P_x = P_x^{NA} + \eta \kappa_1 |\alpha|^2 T
%\kappa_{eff}(|\alpha|^2, \kappa_1, K(\kappa_2)) T 
,\\
\end{array}
\label{eq:gate_error_with_photon_loss}
\end{equation}
where $\kappa_1$ denotes the photon loss rate and $\beta$ is a constant depending on the gate ($\beta = 1$ for Z rotation and $\beta = 2$ for ZZ rotation and CX gate). 
The first terms $P_z^{NA}$ and $P_x^{NA}$ are the non-adiabatic errors due to finite gate time, which become negligibly small for $T \gg 1 /|\Delta(\Delta_d)|$, while the second terms are photon-loss-induced errors which accumulate in time. The loss-induced bit flip is negligible in the fast-gate regime we consider since the induced bit-flip rate is suppressed by a factor $\eta = \eta[|\alpha|^2, \kappa_1/K(\kappa_2)]$, which decreases exponentially for large $|\alpha|^2$ \cite{Lescanne20, Guillaud_Mirrahimi_2019}, and thereby $P_x$ is dominantly given by $P_x^{NA}$.
In this work, we choose the size of the cat as $\alpha^2 = 8$, which is experimentally feasible \cite{Grimm20} and also provides us sufficient noise bias \cite{Puri20, Chamberland_2020}.
There are \LJ{two figures of merit} for the BP gates - the gate fidelity, which is determined by $P_z$, and the noise bias, which is limited by $P_x$. Due to the large loss-induced phase flip rate, fast gates are desirable to obtain high gate fidelity. However, the exponential noise bias might break down in the fast-gate regime due to $P_x^{NA}$. Therefore, it is important to design fast BP gates with suppressed $P_z^{NA}$ and $P_x^{NA}$ in order to have simultaneous high gate fidelity and high noise bias.

% Thus, the total X error probability ($P_x$) is exponentially small for adiabatic gates.
% Since the cat code does not protect against excitation loss (which mostly causes Z error), the gate infidelity is dominated by Z error, with error probability $\beta \kappa_1 |\alpha|^2 T$. 
% To further improve the gate performance, we need to reduce the gate time. 
% However, the exponential suppression of non-adiabatic errors is no longer guaranteed in the fast-gate regime. Therefore, we need to carefully design fast gates, while still preserving the exponential suppression of non-adiabatic errors for cat codes.

First, we would like to point out a fundamental difference in the non-adiabatic errors between Kerr gates and dissipative gates. 
%for Kerr gates and dissipative gates, although they are both limited by non-adiabiticity, their non-adiabatic errors are, by nature, different. 
For dissipative gates, the non-adiabatic errors are associated with accumulated dissipation with continuous information leakage into the environment, which is difficult to restore. For example, as shown in Fig.~\ref{fig:double_well_potential}(d), the linear drive $\Omega(t) a + \Omega^*(t) a^{\dagger}$, which implements the Z rotation, can create leakage outside the cat subspace, which becomes continuous phase flips \cite{Chamberland_2020} when brought back by the engineered two-photon dissipation. In contrast, for Kerr gates shown in Fig.~\ref{fig:double_well_potential}(c), the non-adiabatic errors are coherent off-resonant leakage errors to excited states, which \LJ{might} be reliably eliminated using additional leakage-suppression techniques. %\LJ{Hence, the Kerr gates might work even beyond adiabatic evolution.}

%The Kerr Z-rotation gate originally proposed in \cite{Puri20} uses a simple hard (square) driving pulse. 

%In comparison, we use ``Hard" to refer to the control over Kerr gates in \cite{Puri20} using hard pulses without corrections and use ``Dissipative" to refer to the control over dissipative gates in \cite{Guillaud_Mirrahimi_2019}. 

\begin{figure}[t]
    \centering
    \includegraphics[width = 0.5\textwidth]{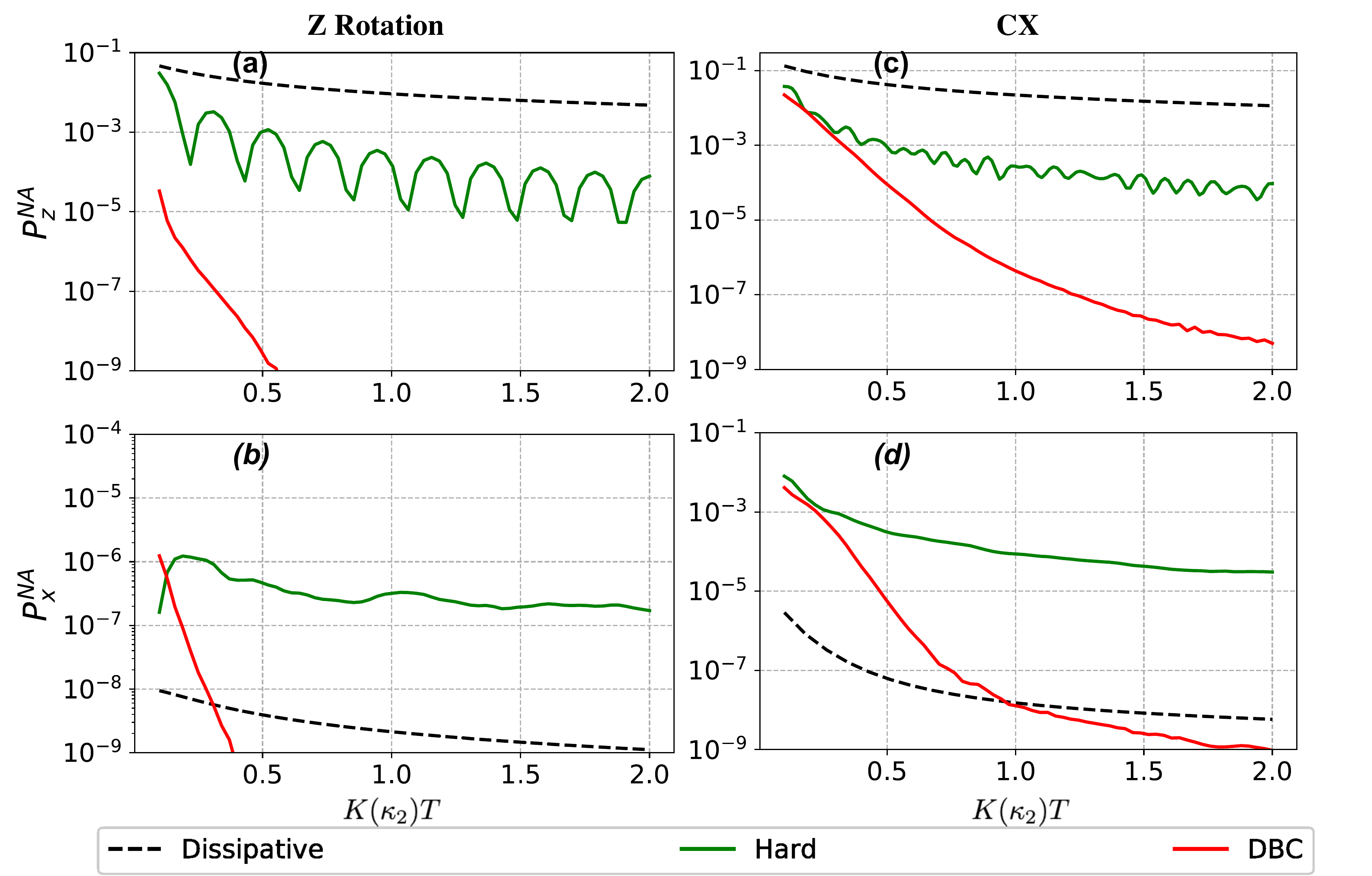}
    \caption{The non-adiabatic Z ($P_{Z}^{NA}$) and X ($P_{X}^{NA}$) errors of Z rotation (a)(b) and CX gate (c)(d) with dissipative gates (black dashed curves), Kerr gates with hard square pulses (green curves) and Kerr gates with DBC control (red curves). The gate time is in units of $K (or \kappa_2)$. Both types of Kerr gates have smaller $P_{Z}^{NA}$ error than the dissipative gates. The DBC control can further suppress $P_{X}^{NA}$ error. The size of the cat is $\alpha^2 = 8$.}
%    with different control schemes. : Z rotation. (c)(d): CX gate.     The dashed black curves are for dissipative gates; the solid green and red curves are for Kerr gates with Hard and DBC control, respectively. }
    \label{fig:CX_noiseless_error_rates}
\end{figure}

As shown in Fig.~\ref{fig:CX_noiseless_error_rates}, the Kerr gates with hard square pulses (green curves, denoted as ``Hard") \cite{Puri20} have smaller $P_z^{NA}$ but larger $P_x^{NA}$ errors than the corresponding dissipative gates (black dashed curves, denoted as ``Dissipative") \cite{Guillaud_Mirrahimi_2019}.
The larger $P_x^{NA}$ error is due to the extra leakage induced by the hard pulse, which leads to coherent tunneling between two wells of the KPO through high excited levels. These numerical results manifest the importance for finer control on the Kerr gates to further suppress the leakage, which motivates our work. Although numerical quantum optimal control method \cite{pinch1995optimal} can be applied to optimizing the gates, analytical solutions which can produce smooth and robust pulses and avoid large-scale numerical optimizations for large cats are desirable. We use derivative-based transition suppression technique \cite{Motzoi_Wilhelm_2013}, which is a variant of the derivative removal by adiabatic gate (DRAG) technique and closely related to the idea of shortcut to adiabaticity (STA) \cite{Guery-Odelin_2019, Theis_Motzoi_Machnes_Wilhelm_2018, Motzoi_Gambetta_Rebentrost_Wilhelm_2009, Motzoi_Wilhelm_2013, Ribeiro_Baksic_Clerk_2017}, to suppress the leakage (diabatic transitions) of the BP Kerr gates so that they can simultaneously have high gate fidelity and high noise bias in the presence of photon loss.

\textit{Derivative-based control of BP Kerr gates} - To suppress the diabatic transitions, We first replace the hard square pulses with a family of truncated Gaussian pulses because of their better frequency selectivity and smoothness \cite{Motzoi_Wilhelm_2013}:
\begin{equation}
\Omega_{G, m}(t)=A_m\left\{\exp \left[-\frac{(t-T / 2)^{2}}{2 \sigma^{2}}\right]-\exp \left[-\frac{(T / 2)^{2}}{2 \sigma^{2}}\right]\right\}^{m}
\end{equation}
where $m$ is chosen such that \LJ{all of the} first $m - 1$ derivatives of $\Omega_{G,m}$ start and end at 0, $A_m$ is a normalization constant and $\sigma$ is chose to be equal to $T$ in this work. Then we add some derivative-based corrections (DBC) $\hat{H}_{DBC}$ to each gate to further suppress the leakage. 
%We use shorthand ``DBC" to refer to this fine control scheme. 
%In comparison, we use ``Hard" to refer to the control over Kerr gates in \cite{Puri20} using hard pulses without corrections and use ``Dissipative" to refer to the control over dissipative gates in \cite{Guillaud_Mirrahimi_2019}. 
%In this work we fix the size of the cat as $\alpha^2 = 8$. 

We summarize our design of DBC control Hamiltonians %for Z rotation, ZZ rotation and CX gate 
as the following (see \cite{SM} for details): 
(1) For the Z rotation, the derivative-based correction simply shapes the driving pulse of the linear drive, $\hat{H}_{DBC} = u(t) \hat{a}^{\dagger} + u^*(t) a$, where 
\begin{equation}
    u(t) = -i \dot{\Omega}_{0}\left(\frac{1}{\Delta_{1}}+\frac{1}{\Delta_{2}}\right) - \frac{\ddot{\Omega}_0}{\Delta_1 \Delta_2} + 0.07 \frac{\Omega_0^3(t)}{\Delta_1^2}
\end{equation}
and the base driving pulse is $\Omega_0(t) = \Omega_{G,2}(t)$.
(2) For the ZZ rotation, in contrast to \cite{Puri20}, we use two-mode squeezing instead of beamsplitter coupling to generate the ZZ interaction between two modes. Similar to the Z rotation, the derivative-based correction amounts to simply shaping the driving pulse of the two-mode squeezing.
(3) For the CX gate, in addition \LJ{to} the original Hamiltonian in \cite{Puri20}, we apply derivative-based corrections that include the off-phase, derivative correction of the original compensation Hamiltonian and a set of single-photon and two-photon drives on the control and target modes. We note that the highest order of nonlinearity of the derivative-based corrections is the same (cubic) as the original Hamiltonian.

%The details and derivations of the DBC control Hamiltonians of the above gates can be found in \cite{SM}.

\textit{Non-adiabatic errors} - 
To evaluate the gate performance with the DBC control, we first numerically obtain the non-adiabatic Z and X error probabilities $P_z^{NA}, P_x^{NA}$ of each gate.
The results for the Z rotation and CX gate, in comparison with those obtained using other control schemes, are shown in Fig.~\ref{fig:CX_noiseless_error_rates} and the non-adiabatic errors of the ZZ rotation are similar to the Z rotation. Overall, both the non-adiabatic Z and X errors of the Kerr gates with DBC control are significantly reduced since they scale with the gate time more favorably. We explicitly provide the scaling of $P_z^{NA}$ with the gate time below since they are easy to analyze. For dissipative gates, $P_z^{NA}$ is given by the integration the induced phase flip rate over gate time and scales \LJ{only} \textit{linearly} with $1/\kappa_2 T$. For Kerr gates with Hard control, the $P_z^{NA}$ is proportional to the fourier component of the hard driving pulse at the energy gap $\Delta$: $|\int_0^T \Omega_0(t) e^{-i\Delta t} dt|^2$, which scales \textit{quadratically} with $1/KT$ (neglecting the fast oscillating terms). And using Gaussian pulses with DBC we can suppress the excitation at $\Delta$ and obtain approximately exponential error scaling with gate time in the fast-gate regime\cite{SM}. 

\begin{figure}[t]
    \centering
    \includegraphics[width = 0.5\textwidth]{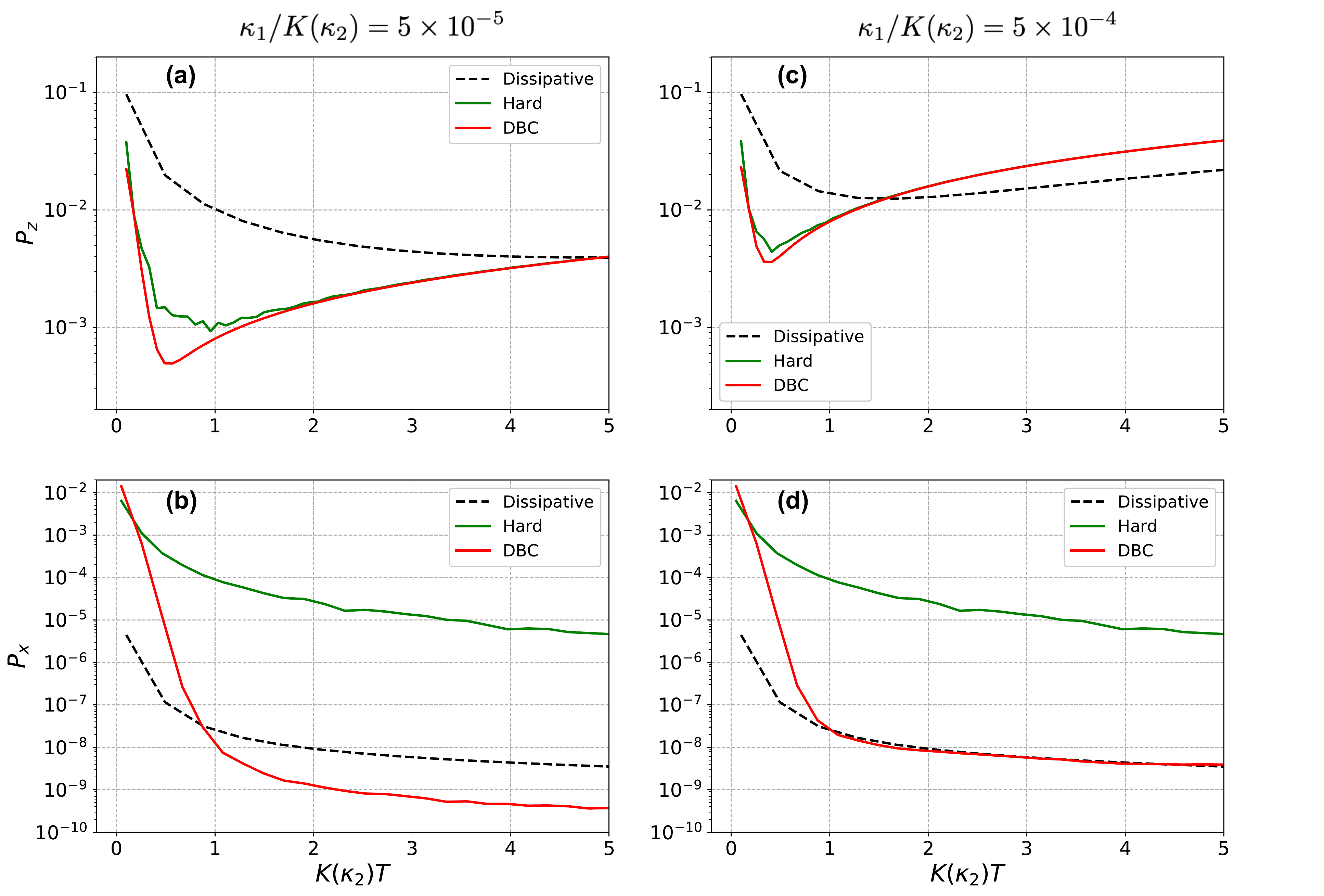}
    \caption{The total Z and X error probabilities for CX gate using different control schemes (same color scheme as Fig.~\ref{fig:CX_noiseless_error_rates}) in the presence of photon loss. (a)(b): $\kappa_1/K(\kappa_2) = 5 \times 10^{-5}$. 
    (c)(d): $\kappa_1/K(\kappa_2) = 5 \times 10^{-4}$. } 
    \label{fig:CX_error_noisy}
\end{figure}

\textit{Gate performance in the presence of photon loss} - 
In the presence of photon loss, the total Z error probability (Eq.~\ref{eq:gate_error_with_photon_loss}) will have a significant new contribution ($\sim \kappa_1 |\alpha|^2 T$), associated with the loss-induced parity change. In contrast, the total X error probability is still dominant by the non-adiabatic error $P_x^{NA}$. In Fig.~\ref{fig:CX_error_noisy}, we numerically obtain Z and X error probability of CX gates at different gate time with $\kappa_1/K(\kappa_2) = 5\times 10^{-5}$ and $5 \times 10^{-4}$.

The total Z error probability $P_z$ determines the gate fidelity. According to Eq.~\ref{eq:gate_error_with_photon_loss}, the gate time can be optimized to minimize the total Z error probability. The Kerr gate can be implemented faster with higher gate fidelity than the dissipative gate and compared to using Hard control, using DBC control can further speed up the gate and improve the gate fidelity. Plugging in the scaling of $P_z^{NA}$ with $T$ we can obtain the scaling of the minimal Z error probability $P_z^*$ with $\kappa_1/K(\kappa_2)$. For dissipative gates, $P_{z, dissi}^* \propto (\frac{\kappa_1}{\kappa_2})^{1/2}$. For Kerr gates with Hard control, $P^*_{z, Hard} \propto (\frac{\kappa_1}{K})^{2/3}$. While for Kerr gates with DBC control, $P_{z, DBC}^*$ can approach the most favorable linear scaling, i.e. $P_{z, DBC}^{*} \propto \frac{\kappa_1}{K}$. We obtain consistent scaling between $P^*_z$ and $\kappa_1/K(\kappa_2)$ based on numerical fits for the CX gates \cite{SM}. Given the same small parameter $\kappa_1/K(\kappa_2)$, our fine control scheme can provide a smaller Z error probability and thus a more favorable optimal gate fidelity.

The total X error probability $P_x$ limits the noise bias of the gates. Compared to the dissipative gate, the $P_x$ of the Kerr gate with Hard control is too high (at reasonable gate time). In contrast, the Kerr CX gate with DBC control can have $P_x$ comparable to or even below that of the dissipative gate. But to obtain larger noise bias we need to use longer gate time than that minimizes $P_z$ and thus compromise a bit with the gate fidelity. 

We stress that the improvement of the BP CX gate using DBC control manifests in two aspects: in the fast-gate regime ($T \sim 1/K$), we can reach much higher gate fidelity due to suppressed $P_z^{NA}$ while maintaining similar noise bias compared to the dissipative gate; in the slower-gate regime, although the advantage in gate fidelity degrades as loss-induced phase flip becomes dominant, the noise bias can be higher due to suppressed $P_x^{NA}$ provided that $\kappa_1/K$ is small. When concatenating the cat code with a second level of code, we should consider the trade-off between the gate fidelity and noise bias and optimize the BP CX gate time to give the best logical performance of the concatenated code.

\textit{Concatenated quantum error correction} - We now compare the performance of different schemes of BP gates in terms of the logical error rates in concatenated QEC. We consider the concatenation of the stabilized cats with a repetition code and use a circuit-level noise model, which includes state preparation errors, idling errors, CX gate errors and measurement errors. 
%See \cite{SM} for more details about our noise model. 
For a distance-$d$ repetition code we repeat the syndrome extraction $d$ times followed by one round of perfect syndrome extraction and decode the full error syndromes using a minimum weight perfect matching (MWPM) decoder for our noise-biased system \cite{SM}.

\begin{figure}[t]
    \centering
    \includegraphics[width = 0.45\textwidth]{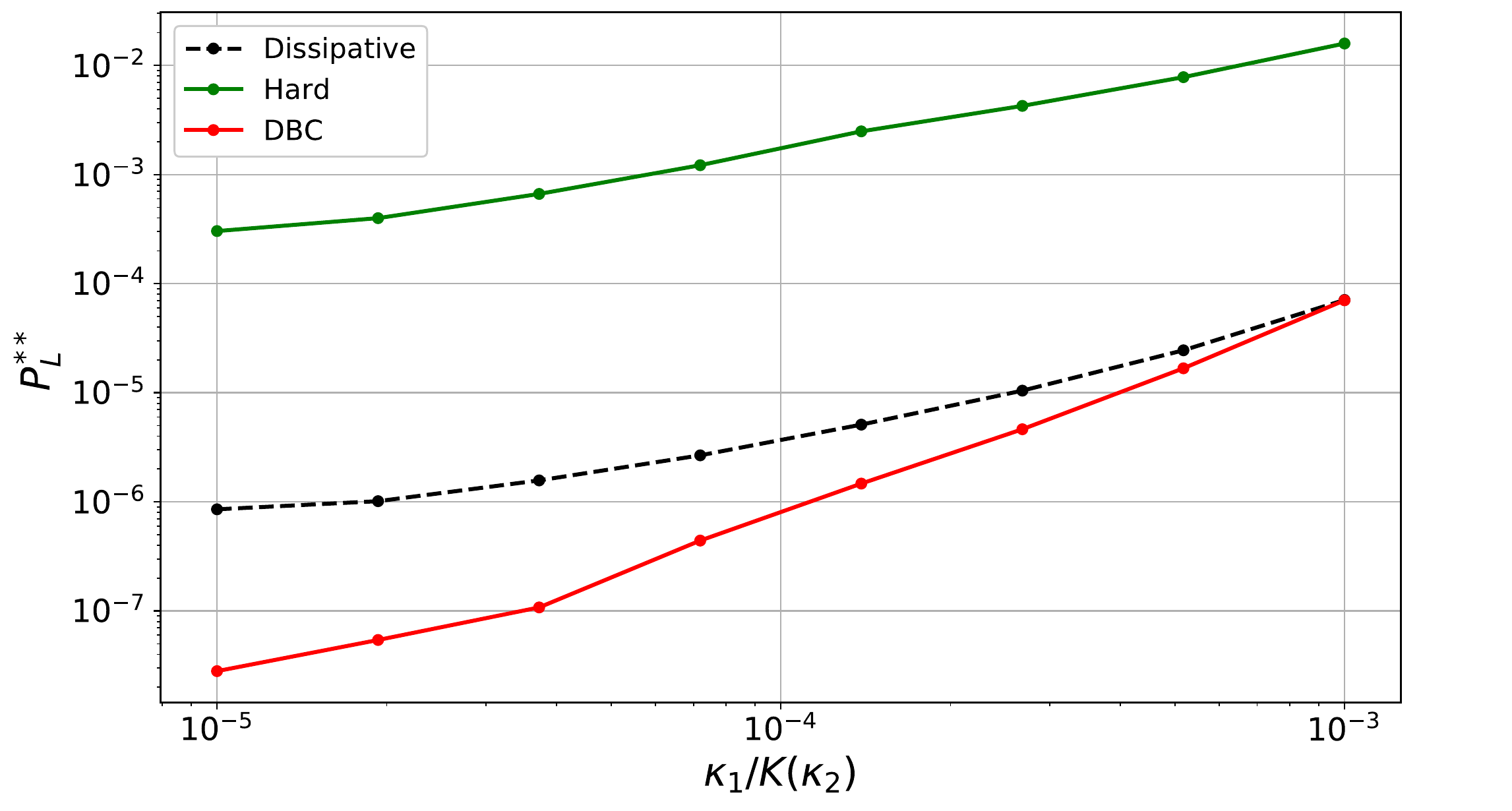}
    \caption{The minimal gate error $P^{**}_L$ of the logical CX gate using physical BP CX gates with different controls.}
    \label{fig:Logical_error_optimal}
\end{figure}

We consider the total logical error rate of a transversal logical CX gate, which is a function of dimensionless parameters of the photon loss rate $\frac{\kappa_1}{K(\kappa_2)}$, the CX gate time $K(\kappa_2) T_{cx}$, and the repetition code distance $d$. Given $\kappa_1$, we can obtain the minimum logical CX gate error rate achievable by the repetition-cat by optimizing $T_{cx}$ and $d$ \cite{SM}:
\begin{equation}
  P^{**}_L(\frac{\kappa_1}{K(\kappa_2)}) = \min_ {T_{cx}, d} P_L(\frac{\kappa_1}{K(\kappa_2)}, K(\kappa_2)T_{cx}, d).
\end{equation}
In Fig.~\ref{fig:Logical_error_optimal}, we plot $P^{**}_L$ as a function of photon loss rate when different physical CX gates are used. Using the Kerr CX gate with DBC control can lead to lower logical gate error compared to using the dissipative CX gate for the range of $\kappa_1/K(\kappa_2)$ we consider. In dramatic contrast, if we use the Kerr CX gate with simple hard pulses the logical gate error is much higher due to lower noise bias. 

In the above we optimize the logical error rates since the repetition-cat cannot suppress logical errors to be arbitrarily small in the regime of finite noise bias. However, we should instead optimize the resource overhead required for reaching a target logical error rate when we consider other concatenation QEC schemes that can arbitrarily suppress logical errors, such as the surface-cat considered in \cite{Chamberland_2020, darmawan2021practical}. In this case, using the fast Kerr gates presented in this work can reduce the resource overhead towards fault tolerance since the dominant phase-flip errors are suppressed to be far below the threshold.

\textit{Discussions} - 
So far, the experimental Kerr parametric nonlinear oscillator can achieve $\kappa_1/K \approx 10^{-3}$ \cite{Grimm20}, which is slightly more favorable than the engineered two-photon dissipation with $\kappa_1/\kappa_2 \approx 10^{-2}$ \cite{Touzard18}, partly because Kerr nonlinearity is less complicated to implement than the two-photon dissipation. 
Note that the single photon loss rate is fairly high in all these experiments \cite{Grimm20, Touzard18, Lescanne20}, to be further reduced in future devices. We expect that $\kappa_1/K(\kappa_2)\le 10^{-4}$ can be achieved \cite{Puri20, Chamberland_2020}, which will enable us to achieve high fidelity logical gates. Based on the improvement of BP gates on Kerr cat in this work, we envision to use an experimental architecture that hybrids the Kerr nonlinearity with tailored dissipation to better protect the cat qubit, which will be investigated in our future work.

\textit{Conclusion} - Compared to the BP dissipative gates, the originally proposed BP Kerr gates using simple hard pulses can be of high gate fidelity yet less bias-preserving since their non-adiabatic Z errors are smaller while X errors are larger. We use a derivative-based transition suppression technique to suppress the leakage of the BP gates on Kerr cat so that both the non-adiabatic Z and X errors can be reduced dramatically. In the presence of photon loss, we show that the Kerr gates with our designed DBC control can have higher gate fidelity while maintaining high noise bias. The improved gate, when applied in concatenated QEC, can lead to lower logical error rates and/or lower resource overhead.

\begin{acknowledgments}
We thank Aashish Clerk, Kyungjoo Noh, Shruti Puri, Harry Putterman and Hugo Ribeiro for help discussions. We acknowledge support from the ARO (W911NF-18-1-0020, W911NF-18-1-0212), ARO MURI (W911NF-16-1-0349), AFOSR MURI (FA9550-19-1-0399), NSF (EFMA-1640959, OMA-1936118, EEC-1941583), NTT Research, and the Packard Foundation (2013-39273). 
\end{acknowledgments}

% \bibliographystyle{aipauth4-1}
% \bibliography{ref}

%merlin.mbs aipauth4-1.bst 2010-07-25 4.21a (PWD, AO, DPC) hacked
%Control: author (9) reversed initials

%\bibliographystyle{apsrev4-2}
%\bibliography{ref} 
%\bibliographystyle{aipauth4-1}
%\bibliographystyle{unsrtnat}


\end{document}

%\preprint{}
%Title of paper$\epsilon_{G,m}^{(1)}(T)$
\title{Supplemental Materials}
\maketitle

\section{The shifted Fock basis and the Kerr-cat eigenbasis}
Simulating a large cat qubit using the usual fock basis is inefficient due to the wide photon number distribution of a large coherent state. In contrast, one can work with the so called shifted Fock basis, as proposed in Ref.~\cite{Chamberland_2020}, to simplify the analysis since typically only the first few excited states are populated. The shifted basis is defined as:
\begin{equation}
|\phi_n, \pm \rangle \equiv \mathcal{N}_{n,\pm]} [\hat{D}(\alpha) \pm (-1)^n \hat{D}(- \alpha)] |n \rangle,
\end{equation}
where $\hat{D}(\alpha)$ is the displacement operator and $\mathcal{N}_{n,\pm}$ is the normalization constant. In this basis, the Hilbert space is split into two subspaces, labeled by the photon number parity $+$ (even) and $-$ (odd), respectively. Thus, we can effectively represent each shifted Fock state as a tensor product of a ``parity qubit" labeling the parity and a Fock state labeling the excitation level:
\begin{equation}
|p\rangle \otimes |n'\rangle \equiv |\phi_n, p \rangle .
\end{equation}
Since the logical information of the cat qubit is encoded in the parity, this ``parity qubit" can also be viewed as a ``logical qubit" carrying one bit of logical information of the cat qubit with Z-basis states $|0\rangle = \frac{1}{\sqrt{2}}(|+\rangle + |-\rangle)$ and $|1\rangle = \frac{1}{\sqrt{2}}(|+\rangle - |-\rangle)$. We note that these shifted Fock states are not exactly mutual orthogonal. But they are nearly orthogonal for $n < |\alpha|^2/4$. We neglect the non-orthogonality for now when analyzing the low excited states. In this shifted fock basis, the annilation operator can be expressed as:
\begin{equation}
\hat{a} = \hat{Z} \otimes (\hat{a}' + \alpha),
\end{equation}
where $\hat{Z}$ flips the phase of the "logical qubit" (or flips the parity of the "parity qubit") and $\hat{a}' \equiv \sum_n' \sqrt{n'} |n' - 1\rangle \langle n'|$ is the bosonic annilation operator defined on the exication level of the shifted fock states. Based on this representation of $\hat{a}$ we can write the Hamiltonian of a Kerr-cat qubit (Kerr parametric oscillator):
\begin{equation}
\begin{aligned}
H_{KPO} &=-K \left(a^{2 \dagger}-\alpha^{2}\right)\left(a^{2}-\alpha^{2}\right) \\
&=-K I \otimes \left[4 \alpha^{2} a^{\prime \dagger} a^{\prime}+2 \alpha\left(a^{\prime 2 \dagger} a^{\prime}+a^{\prime \dagger} a^{\prime 2}\right)+a^{\prime 2 \dagger} a^{\prime 2}\right].
\end{aligned}
\end{equation}

In the large $\alpha$ limit $H_{KPO}$ is dominant by $- 4\alpha^2 K I \otimes a^{\prime \dagger} a^{\prime}$, which is diagonal in the shifted Fock basis. However, the off-diagonal elements can not be neglected in this work and they have a perturbative effects on the eigenvalues with respect to $1/\alpha$. Since the Kerr Hamiltonian preserves the photon number parity we can express its eigenstates as $|\psi_n, p\rangle = |p\rangle \otimes |n''\rangle$. We can perturbatively calculate $|n''\rangle$ in the basis of $\{|n'\rangle \}$ to the first order of $\frac{1}{\alpha}$:
% \begin{equation}
% \begin{array}{c}
% \left|0^{\prime \prime}\right\rangle=\left|0^{\prime}\right\rangle \\
% \left|1^{\prime \prime}\right\rangle=\left|1^{\prime}\right\rangle-\frac{\sqrt{2} \alpha}{2 \alpha^2 + 1}\left|2^{\prime}\right\rangle+O\left(\left(\frac{1}{\alpha}\right)^{2}\right) \\
% \left|2^{\prime \prime}\right\rangle=\left|2^{\prime}\right\rangle+\frac{\sqrt{2}\alpha}{2 \alpha^2 + 1}\left|1^{\prime}\right\rangle-\frac{\sqrt{3}\alpha}{\alpha^2 + 1}\left|3^{\prime}\right\rangle+O\left(\left(\frac{1}{\alpha}\right)^{2}\right).
% \end{array}
% \end{equation}
\begin{equation}
    \left|n^{\prime \prime}\right\rangle=\frac{\alpha(n-1) \sqrt{n}}{2 \alpha^{2}+(n-1)}|n-1'\rangle +|n'\rangle-\frac{\alpha n \sqrt{n+1}}{2 \alpha^{2}+n}|n+1'\rangle+O\left((1 / \alpha)^{2}\right),
\end{equation}
where $n \geq 1$ and $|0''\rangle = |0'\rangle$. 

% The eigenenergies can also by perturbatively calculated:
% \begin{equation}
% E_n = - K_h [4\alpha^2  n + n(n - 1)] + O((\frac{1}{\alpha})^2)
% \end{equation}

If we only consider the first three pair of eigenstates, i.e., $n' \leq 2$, we can express $\hat{a}'$ as:
% \begin{equation}
% \hat{a}' = \sigma_{0,1}^{-}+\sqrt{2}\sigma_{1,2}^{-}-\frac{2\alpha}{2\alpha^2 + 1} \Pi_{1}-\frac{3\alpha}{\alpha^2 + 1} \Pi_{2}+\frac{\sqrt{2} \alpha}{2 \alpha^2 + 1} \sigma_{0,2}^{-}+O\left(\left(\frac{1}{\alpha}\right)^{2}\right)
% \end{equation}

\begin{equation}
\hat{a}' = \sigma_{0,1}^{-}+\sqrt{2}\sigma_{1,2}^{-}-\lambda_1 \Pi_{1}-\lambda_2 \Pi_{2}+ \eta \sigma_{0,2}^{-},
\end{equation}
where $\lambda_1, \lambda_2, \eta = \frac{2\alpha}{2\alpha^2 + 1}, \frac{3\alpha}{\alpha^2 + 1}, \frac{\sqrt{2} \alpha}{2 \alpha^2 + 1} + O\left(\left(\frac{1}{\alpha}\right)^{2}\right)$, with reduced Pauli operators and projectors in the Kerr-cat eigenbasis:
\begin{equation}
\begin{array}{c}
\sigma^-_{i,j} \equiv |i''\rangle \langle j''|, \\
\Pi_i \equiv |i''\rangle \langle i''|. \\
\end{array}
\end{equation}
$\lambda_1, \lambda_2, \eta$ can be calculated more accurately by adding higher-order corrections. For $\alpha = \sqrt{8}$ used in this work, we can numerically obtain:
\begin{equation}
\lambda_1 = 0.4, \lambda_2 = 1.08, \eta = 0.256.
\end{equation}

\section{Estimation of off-resonant excitation in the asymptotic limit via Fourier analysis \label{sec:fourier_analysis}}
In this section, following \cite{Motzoi_Wilhelm_2013} we provide the estimation of off-resonant excitations in the asymptotic weak-drive limit, based on which we define the order of different transition elements in a Hamiltonian.  In the case of constant energy gap, an off-resonant excitation can be estimated via the Fourier spectrum of its driving pulse in the asymptotic weak-drive limit. Only considering a single transition element $\hat{h}_k \equiv |\psi_k^{to}\rangle \langle \psi_k^{from}|$ associated with constant energy gap $\Delta_k$ that is driven by $\Omega(t)$, the propagator in the interaction picture is give by:
\begin{equation}
\hat{U}_I(t) = \mathcal{T} \exp[-i\int_0^{t} dt' (\Omega(t') \hat{h}_k e^{i\Delta_k t'} + h.c.)],
\end{equation}
where $\mathcal{T}$ is the time-ordering operator. In the weak-drive limit, $U_I(t)$ is dominantly given by the first-order Dyson expansion:

\begin{equation}
\hat{U}_I^{(1)}(t) = -i \int_0^{t} dt' [\Omega(t') \hat{h}_k e^{i\Delta_k t'} + h.c.],
\end{equation}
then at certain time $T$ the off-resonant transition strength is given by the finite-time Fourier transform of $\Omega(t)$:
\begin{equation}
\langle \psi_k^{to}| \hat{U}_I^{(1)}(T) |\psi_k^{from}\rangle = \mathcal{F}(\Omega, \Delta_k, T) \equiv \int_0^T \Omega(t) e^{-i\Delta_k t} dt,
\end{equation}
and the population in the excited state is given by $|\mathcal{F}(\Omega, \Delta_k, t)|^2$. This finite-time Fourier transform can be connected to the standard Fourier transform by assuming $\Omega(t)$ is truncated outside the [0,T] time window or $\Omega(t)$ smoothly vanishes outside [0,T], which is usually the case for a gate pulse that starts and ends at 0.

For a pulse whose time derivatives also start and end at 0 up to order m, its Fourier spectrum has the property that:
\begin{equation}
\mathcal{F}(\Omega, \Delta, T) = (-i)^n \mathcal{F}(\frac{\frac{d^n}{dt^n}\Omega(t)}{\Delta^{n}}),
\label{eq:Fourier_property}
\end{equation}
for $n = 1,2,...m + 1$. Moreover, the Fourier spectrum of the product of derivatives of $\Omega$ can be converted to the Fourier spectrum of polynomials of $\Omega$:
\begin{equation}
\mathcal{F}\left((\Omega(t))^{\sum n_{k}}, \Delta, T\right)=\Theta\left(\mathcal{F}\left(\prod_{k}\left(\frac{1}{\Delta^{k}} \frac{d^{k} \Omega(t)}{d t^{k}}\right)^{n_{k}}, \Delta, T\right)\right).
\label{eq:derivative_rule}
\end{equation}

To facilitate the perturbative analysis used throughout this work, we define the order of the off-resonant transition elements with respect to the driving pulse $\Omega(t)$ and the energy gap $\Delta$ in the following way: we define the n-th order transition elements as those which will create n-th order excitation $e^{(n)}$ after gate time $T$:
\begin{equation}
e^{(n)} \propto |\mathcal{F}(\frac{\Omega^n}{\Delta^{n-1}}, \Delta, T)|^2.
\label{eq:error_order}
\end{equation}
Based on this definition the n-th order elements have a coefficient of $\frac{\Omega^n}{\Delta^{n-1}}$ or a product of time derivatives of $\Omega$ according to Eq.~\ref{eq:derivative_rule}. We note that the definition of $e^{(n)}$ coincides with the more conventional definition $e^{(n)} \propto (\frac{\Omega_0}{\Delta})^n$ if $\Omega(t)$ is a hard (square) pulse that has the amplitude $\Omega_0$.

 % This is the classical picture how derivative-based approach, which will be discussed in the next section, can be applied to eliminate off-resonant transition at certain gap frequencies. 

We use Eq.~\ref{eq:error_order} to estimate the non-adiabatic errors of the gates using different control schemes in the main text. We consider two type of driving pulses in this work, one is the hard (square) pulse $\Omega_h(t) = \frac{\pi}{T}$ and the other is the truncated Gaussian pulses:
\begin{equation}
\Omega_{G, m}(t)=A_{m}\left\{\exp \left[-\frac{(t-T / 2)^{2}}{2 \sigma^{2}}\right]-\exp \left[-\frac{(T / 2)^{2}}{2 \sigma^{2}}\right]\right\}^{m},
\end{equation}
where $\sigma$ is set to be $T$.

For gates using hard pulses, the leakage is given by the first-order excitation: 
\begin{equation}
e_{h}^{(1)}(T) \propto |\mathcal{F}(\Omega_h, \Delta, T)|^2 = 4\frac{\sin^2 \Delta T}{(\Delta T)^2},
\label{eq:excitation_hard}
\end{equation}
which scales quadratically with $1/\Delta T$.

\begin{figure}[h!]
    \centering
    \includegraphics[width = 0.5\textwidth]{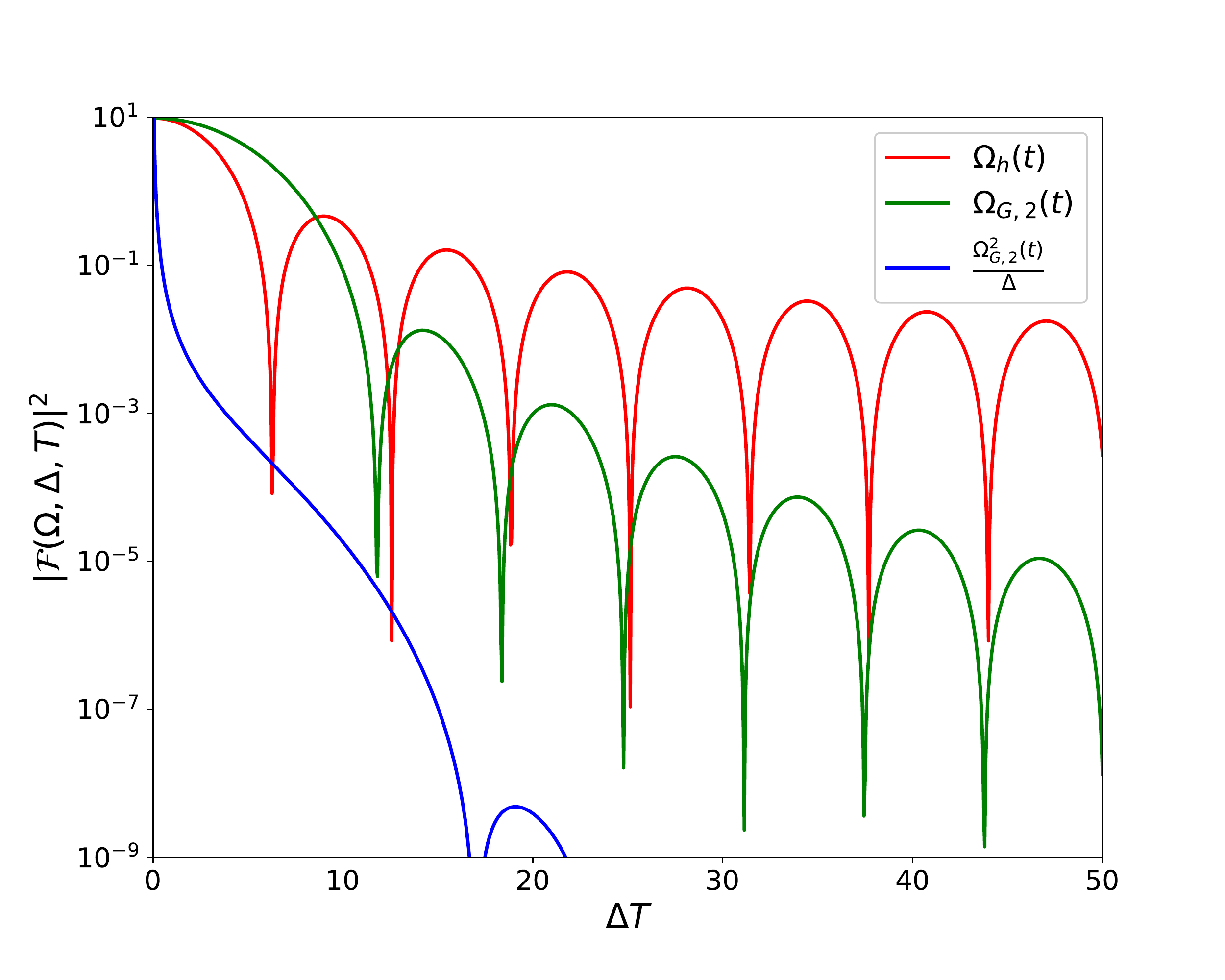}
    \caption{The off-resonant excitation of different pulses, which is estimated by their finite-time fourier spectrum. Red line:  the first-order excitation of the hard pulse Eq.~\ref{eq:excitation_hard}; Green line: the first-order excitation of the truncated Gaussian pulse Eq.~\ref{eq:excitation_Gaussian_1od}; Blue line: the second-order excitation of the truncated Gaussian pulse Eq.~\ref{eq:excitation_Gaussian_2od}. }
    \label{fig:fourier_spectrum}
\end{figure}

If we simply replace the base driving pulse with truncated Gaussian pulses the leakage is then $e^{(1)}_{G, m}$; if we add derivative-based corrections to suppress the first-order excitation (which will be discussed in the next section) the residual leakage is given by the second-order excitation $e^{(2)}_{G,m}$.  $e^{(1)}_{G, m}$ and  $e^{(2)}_{G, m}$ are defined as:
% \begin{equation}
% \begin{aligned}
% e_{G,m}^{(1)}(T) & \propto |\mathcal{F}(\Omega_{G,m}, \Delta, T)|^2 \\
% \ e_{G,m}^{(2)}(T) & \propto |\mathcal{F}(\frac{\Omega^2_{G,m}}{\Delta}, \Delta, T)|^2.\\
% \end{aligned}
% \end{equation}
\begin{equation}
e_{G,m}^{(1)}(T)  \propto |\mathcal{F}(\Omega_{G,m}, \Delta, T)|^2,
\label{eq:excitation_Gaussian_1od}
\end{equation}
\begin{equation}
\ e_{G,m}^{(2)}(T) \propto |\mathcal{F}(\frac{\Omega^2_{G,m}}{\Delta}, \Delta, T)|^2.
\label{eq:excitation_Gaussian_2od}
\end{equation}
Their analytical expressions are lengthy, so we instead numerically plot them in Fig.~\ref{fig:fourier_spectrum}. Compared to the first-order excitation of the hard pulse Eq.~\ref{eq:excitation_hard}, shown by the green line, the first-order excitation of the truncated Gaussian pulse Eq.~\ref{eq:excitation_Gaussian_1od}, shown by the red line, is more narrow-band and has a faster-decaying envelop as $\Delta T$. And going to the second order (Eq.~\ref{eq:excitation_Gaussian_2od}, blue line) gives further dramatic improvement. 

% As shown by the green curve, $\epsilon_{G,m}^{(1)}(T)$  first decreases exponentially with $T$ in the short-T regime as a result of being Gaussian pulse and then scales with T in power law due to the truncation effect in the long-T regime. For $\epsilon_{G,m}^{(2)}(T)$ shown by the blue line, the approximately exponential scaling with T can be extended to the regime where the spectrum weight is sufficiently small.

In addition to facilitating the order counting and scaling analysis, the Fourier analysis also gives us a classical picture of how derivative-based transition suppression works. Using Eq.~\ref{eq:Fourier_property}, for an arbitrarily smooth driving pulse $\Omega(t)$, one can add its higher-order derivatives $\{\frac{d^n}{dt^n}\Omega(t)\}$ to create a set of ``spectral holes" at one or more gap frequencies. In general, to create N holes at frequencies $\Delta_1, \Delta_2, ... \Delta_N$, one can modify $\Omega(t)$ as\cite{Motzoi_Wilhelm_2013}:
\begin{equation}
\Omega' = \Omega- i \sum_{k} \frac{\partial_{t} \Omega}{\Delta_{k}}-\sum_{k} \sum_{j<k} \frac{\partial_{t}^{2} \Omega}{\Delta_{k} \Delta_{j}}+\ldots+\frac{(-i)^{N} \partial_{t}^{N} \Omega}{\Delta_{1} \Delta_{2} \cdots \Delta_{N}}, 
\end{equation}
provided that the first $N - 1$ derivatives of $\Omega(t)$ start and end at 0.

\section{Derivative-based corrections to the BP gates}
In this section we present the details of our designed derivative-based corrections to each BP gate. The corrections are derived using the derivative-based transition suppression technique ~\cite{Theis_Motzoi_Machnes_Wilhelm_2018}\cite{Motzoi_Wilhelm_2013}. 

We first derive the corrections to the Z rotation and ZZ rotation as the corrections to these two gates are simpler and illustrate the core idea. The system Hilbert space can be split into the following form: $\mathcal{H} = \mathcal{H}_{logical} \oplus \mathcal{H}_{leak}$, where $\mathcal{H}_{logical}$ is the logical subspace (cat subspace) inside which we would like our system to stay while $\mathcal{H}_{leak}$ is the subspace which we prevent our system from leaking into. We define $\hat{P}$ as the projector onto $\mathcal{H}_{logical}$ while $\hat{Q}$ as the complementary projector onto $\mathcal{H}_{leak}$. The original Hamiltonian we apply to generate the desired dynamics is typically in the form:
\begin{equation}
\hat{H}_{original} = \hat{H}_0 (t) + \hat{V}(t), 
\end{equation}
where $\hat{H}_0 = 0 \otimes \Delta \sum_k \frac{\Delta_k}{\Delta} \Pi_k$ is diagonal and only has support on the leakage subspace. $\hat{V}(t) = \Omega_0(t) \hat{V}_0^+ + h.c.$ is the control Hamiltonian whose projected part $\hat{P}\hat{V}(t)\hat{P}$ generates the desired logical operation while block off-diagonal part $\hat{P}\hat{V}(t)\hat{Q} + \hat{Q}\hat{V}(t)\hat{P}$ causes leakage. Specifically, $\hat{Q} \hat{V}_0^+ \hat{P} = \sum_k \lambda_k \hat{h}_k$ contains diabatic transitions to different leakage level $|k\rangle$ with transition strength $\lambda_k$. These diabatic transitions are off-resonant since $\mathcal{H}_{leak}$ is gapped from $\mathcal{H}_{logical}$ by a set of energy gaps $\{\Delta_k \}$ in $H_0$. The leaked population can be estimated by the finite-time fourier spectrum of $\Omega_0(t)$ at different gap energies in the asymptotic limit (see previous section).

To suppress the leakage, we aim to find a frame transformation, the so-called DRAG frame define by $\hat{D}(t)$, that can block-diagonalize the original Hamiltonian $\hat{H}_{original}(t)$ and obtain the effective Hamiltonian in this DRAG frame:
\begin{equation}
\hat{H}_{eff} = \hat{D}(t) \hat{H}_{original} \hat{D}^{\dagger}(t) + i \dot{\hat{D}}(t) \hat{D}^{\dagger}(t),
\end{equation}
where $\hat{D}(t) \hat{H}_{original} \hat{D}^{\dagger}(t) = \hat{P} \hat{D}(t) \hat{H}_{original} \hat{D}^{\dagger}(t) \hat{P} + \hat{Q} \hat{D}(t) \hat{H}_{original} \hat{D}^{\dagger}(t) \hat{Q}$ is block diagonal and $i \dot{\hat{D}}(t) \hat{D}^{\dagger}(t)$ in general contains block off-diagonal part. If we are able to remove the diabatic term $i \dot{\hat{D}}(t) \hat{D}^{\dagger}(t)$ by adding some corrections we can remove all the leakage in this DRAG frame. If we further ensures that the DRAG frame coincides with the lab frame at the beginning and the end of the gate, i.e. $\hat{D}(0) = \hat{D}(T) = 0$ where $T$ is the gate time, we then successfully remove all the leakage by the end of the gate in the lab frame. This correction corresponds to modifying the original Hamiltonian by adding a derivative-based correction:
\begin{equation}
\hat{H}_{modified} = \hat{H}_{original} + \hat{H}_{DBC}(t),
\end{equation}
where $\hat{H}_{DBC} = - i \hat{D}^{\dagger}(t) \dot{\hat{D}}(t)$. However, in practice we can neither perfectly block-diagonalize $\hat{H}_{original}(t)$ nor perfectly apply the desired corrections. So we have to perturbatively obtain both the frame transformation $\hat{D}(t)$ and the correction $\hat{H}_{DBC}$.

% Here we specify the form $\hat{H}_0$ and $\hat{V}$ that we deal with in this work:
% \begin{equation}
% \begin{array}{c}
% \hat{H}_0 = \sum_{|m\rangle \in \mathcal{H}_{leakage}} \Delta_m |m\rangle \langle m| \\
% \hat{V}(t) = \Omega_0(t) \hat{V}_0^+ + h.c. \\
% \hat{V}_0^+ = \sum_{|i\rangle, |j\rangle \in \mathcal{H}_{logical}} \lambda_{ij} |i\rangle \langle j| +  \sum_{|m\rangle, |n\rangle \in \mathcal{H}_{leakage}} \lambda_{mn} |m\rangle \langle n| +  \sum_{|i\rangle \in \mathcal{H}_{logical}, |m\rangle \in \mathcal{H}_{leakage}} \lambda_{mi} |m\rangle \langle i|
% \end{array}
% \end{equation}
We define the adiabatic parameter as 
$\epsilon \equiv |\Omega_0/\Delta|$, upon which we perform the perturbation expansion. Specifically, we define $\hat{D}(t) = \exp[i \hat{S}(t)]$, where $\hat{S}(t) = \sum_{j=1} \hat{S}^{(j)}$ is expanded to different orders in $\epsilon$.  Similarly, $\hat{H}_{DBC} = \sum_j \hat{H}^{(j)}_{DBC}$ is also expanded to different orders in $\epsilon$. We note the order the derivatives of $\Omega_0(t)$ should be counted by the rule $\frac{\partial^k_t \Omega_0}{\Delta^{k+1}} = \Theta(\epsilon)$. The effective Hamiltonian in this DRAG frame can also be expanded to different orders in $\epsilon$ using the time-dependent Schrieffer-Wolff (SW) expansion:
\begin{equation}
\begin{aligned}
\hat{H}_{eff} & = \hat{D}(t) \hat{H}_{modified} \hat{D}^{\dagger}(t) + i \dot{\hat{D}}(t) \hat{D}^{\dagger}(t) \\
& = \sum_{n} \frac{1}{n !}[\hat{H}, - i \hat{S}]_{n} + (-i)^n \sum_{n} \frac{1}{(n+1) !}[\dot{\hat{S}}, \hat{S}]_{n} \\
& = \sum_{j = 0} \hat{H}^{(j)}_{eff}(t),
\end{aligned}
\end{equation}
where $[\hat{A}, \hat{B}]_{n}=\left[[\hat{A}, \hat{B}]_{n-1}, \hat{B}\right] \text { and }[\hat{A}, \hat{B}]_{0}=\hat{A}.$ We note that $\hat{H}_0$ is of order 0 and $\hat{V}(t)$ is of order 1. We explicitly list $\hat{H}_{eff}$ up to the second order below:
\begin{equation}
\begin{array}{c}
\hat{H}_{eff}^{(0)} = \hat{H}_0 \\
\hat{H}_{eff}^{(1)} = \hat{V}+i\left[\hat{S}^{(1)}, \hat{H}_{0}\right] + \hat{H}_{DBC}^{(1)} + \dot{\hat{S}}^{(1)} \\
\hat{H}_{eff}^{(2)}=  i\left[\hat{S}^{(2)}, \hat{H}_0\right] + i\left[\hat{S}^{(1)}, \hat{V} + \hat{H}_{DBC}^{(1)}\right]-\frac{1}{2}\left[\hat{S}^{(1)},\left[\hat{S}^{(1)}, \hat{H}_{0}\right]\right] + \hat{H}_{DBC}^{(2)} + \dot{\hat{S}}^{(2)} - i \left[\dot{\hat{S}}^{(1)}, \hat{S}^{(1)}\right].
\end{array}
\label{eq:SW_expansion}
\end{equation}

In this work, we only correct the leakage error to the first few excited states to the first order, i.e. $Tr[{H^{(1)}_{eff}} \hat{h}_k] = 0$ for $k = 1,2,...,N$, which can be satisfied simply by shaping the control pulse:
\begin{equation}
H_{DBC}^{(1)}(t) = u(t) \hat{V}_0^+ + h.c.,
\end{equation}
where $u(t)$ is the classical solution that corresponds to creating N ``spectral holes" at N different gap energies $\{\Delta_k \}$\cite{Motzoi_Wilhelm_2013}:
\begin{equation}
u= -i \sum_{k} \frac{\partial_{t} \Omega_0}{\Delta_{k}}-\sum_{k} \sum_{j<k} \frac{\partial_{t}^{2} \Omega_0}{\Delta_{k} \Delta_{j}}+\ldots+\frac{(-i)^{N} \partial_{t}^{N} \Omega_0}{\Delta_{1} \Delta_{2} \cdots \Delta_{N}
\label{eq:clssical_sol_pulse}
}
\end{equation}
and the corresponding first order DRAG frame-transformation is given by:
\begin{equation}
\hat{S}^{(1)} = i\sum_k \frac{\Omega_0}{\Delta_k}\hat{h}_k + \sum_k \sum_{j \neq k}\frac{\partial_t \Omega_0}{\Delta_j \Delta_k} \hat{h}_k - \sum_k \sum_{j \neq k} \sum_{i\neq j, k} \frac{\partial^2_t \Omega_0}{\Delta_i \Delta_j \Delta_k} \hat{h}_k ... + \sum_k \frac{(-1)(-i)^N \partial_t^{N - 1} \Omega_0}{\Delta_1 \Delta_2 ... \Delta_N} \hat{h}_k + h.c.
\label{eq:clssical_sol_trans}
\end{equation}

We note that this frame transformation can be linked to the superadiabatic expansion by applying successive frame transformation defined by $\exp [i \hat{S}^{(1)}_l]$, where $l$ labels the l-th superadiabatic transformation and $S^{(1)}_l$ contains the term in Eq.~\ref{eq:clssical_sol_trans} with l-th order derivative of $\Omega_0(t)$.

By adding $H_{DBC}^{(1)}$ the leakage errors are then suppressed to the second order, i.e. $O(|\mathcal{F}(\frac{\Omega_0^2}{\Delta}, \Delta, T)|^2)$. However, although the leakage error brought by the higher-order expansions in Eq.~\ref{eq:SW_expansion} is smaller than the original first-order leakage, there will be phase or rotation errors acting on the logical subspace directly resulting from those higher-order expansions. As these terms are not associated with any energy gap, their contribution can be comparable or even larger than the residual leakage error. To deal with these errors we need to calculate $\hat{H}_{eff}^{(2)}$ (or higher-order terms) given $\hat{S}^{(1)}$ and $\hat{H}_{DBC}^{(1)}$ and add high order corrections to $\hat{H}_{DBC}$. 

Now we apply the derived correction strategy to the Z rotation and ZZ rotation respectively:

\subsection{Z Rotation}
The original Hamiltonian implementing a Z rotation on single cat in \cite{Puri20} is:
\begin{equation}
\hat{H}_{original}=\hat{H}_{0}+\hat{V}=-K_{h}\left(\hat{a}^{2 \dagger}-\alpha^{2}\right)\left(\hat{a}^{2}-\alpha^{2}\right) + \Omega_{0}(t)\hat{a}^{\dagger}+ \Omega_0^* (t) \hat{a},
\end{equation}
where $\hat{H}_0 = -K_{h}\left(\hat{a}^{2 \dagger}-\alpha^{2}\right)\left(\hat{a}^{2}-\alpha^{2}\right)$, $\hat{V} = \Omega_{0}(t)\hat{a}+ \Omega_0^* (t) \hat{a}^{\dagger}$. $\Omega_0$ is a real hard pulse. 

Working in the eigenbasis of the Kerr-cat Hamiltonian and only considering the first three pair of eigenstates, we can express $\hat{H}_{0}$ and $\hat{V}$ as:
\begin{equation}
	\begin{array}{c}
	\hat{H}_0 = \hat{I} \otimes ( \Delta_1 \hat{\Pi}_1 + \Delta_2 \hat{\Pi}_2)\\
	\hat{V} = 2\alpha \Omega_0(t) \hat{Z} \otimes [\hat{\Pi}_0 + (1 - \lambda_1)\hat{\Pi}_1 + (1 - \lambda_2) \hat{\Pi}_2]  + [ \Omega_0 (t) \hat{Z} \otimes\left(\hat{\sigma}_{0,1}^{+}+\sqrt{2}\hat{\sigma}_{1,2}^{+}+ \eta \hat{\sigma}_{0,2}^{+}\right) + h.c.].
	\end{array}
\end{equation}

In the second term of $\hat{V}(t)$ there are two diabatic transitions $\hat{Z}\otimes \hat{\sigma}^+_{0,1}, \hat{Z}\otimes \hat{\sigma}^+_{0,2}$ associated with two gap frequencies $\Delta_1, \Delta_2$ controlled by the same driving pulse. Then to suppress the first-order leakage we first replace the base driving pulse as the Gaussian pulse with second-order smoothness $\Omega_0(t) = \Omega_{G,2}(t)$, then add $H_{DBC}$ by shaping the base driving pulse, i.e. $\hat{H}_{DBC} = u(t)\hat{a}^{\dagger}+ u^* (t) \hat{a}$. According to Eq.~\ref{eq:clssical_sol_pulse} the first-order correction pulse is:
\begin{equation}
	u^{(1)}(t) = -i \dot{\Omega}_{0}\left(\frac{1}{\Delta_{1}}+\frac{1}{\Delta_{2}}\right) - \frac{\ddot{\Omega}_0}{\Delta_1 \Delta_2}, 
\end{equation}
and the corresponding first-order DRAG transformation (according to Eq.~\ref{eq:clssical_sol_trans}) is:
\begin{equation}
	\hat{S}^{(1)}(t) = i \Omega_0 \hat{Z} \otimes\left(\frac{\hat{\sigma}_{0,1}^{+}}{\Delta_{1}} +\frac{\hat{\sigma}_{1,2}^{+}}{\Delta_{1}} +\eta \frac{\hat{\sigma}_{0,2}^{+}}{\Delta_{2}}\right) + \frac{\dot{\Omega}_{0}}{\Delta_{1} \Delta_{2}} \hat{Z} \otimes\left(\hat{\sigma}_{0,1}^{+}+\hat{\sigma}_{1,2}^{+}+\eta \hat{\sigma}_{0,2}^{+}\right) + h.c.
\end{equation}
 
Then we can block-diagonalize $\hat{H}_{eff}^{(1)}$, i.e., $\hat{P}\hat{H}_{eff}^{(1)}\hat{P} = 2\alpha \Omega_x (t) \hat{Z} \otimes \hat{\Pi}_0$, $\hat{Q}\hat{H}_{eff}^{(1)}\hat{P} = 0$ (only to the first three pair of eigenstates). The residual error is then of order $O(|\mathcal{F}(\frac{\Omega_0^2}{\Delta}, \Delta, T)|^2)$ which comes from $\hat{H}_{eff}^{(2)}$ and higher order expansions. In addition to the leakage error, however, there will be terms from $\hat{H}_{eff}^{(3)}$ (and higher-order expansions) that cause over-rotation ($\hat{Z}\otimes \hat{\Pi}_0$). The over-rotation angle is given by $\delta \theta \propto \mathcal{F}(\frac{\Omega_0^3}{\Delta_1^2}, 0, T)$, which can be corrected by re-normalize the rotation angle or adding addition term to $u$:
\begin{equation}
u(t) = u^{(1)}(t) + c_0 \frac{\Omega_0^3(t)}{\Delta_1^2}.
\end{equation} 
Instead of doing lengthy calculations we simply numerically optimize $c_0$ and obtain $c_0 \approx 0.07$.

\subsection{ZZ rotation}
The original Hamiltonian implementing a ZZ rotation on two cats in \cite{Puri20} is:
\begin{equation}
\begin{array}{c}
\hat{H}_{original}(t)  = \hat{H}_0 + \hat{V}(t)  \\
\hat{H}_0 = - K (\hat{a}_c^{2\dagger} - \alpha^2)(\hat{a}_c^{2} - \alpha^2) - K (\hat{a}_t^{2\dagger} - \alpha^2)(\hat{a}_t^{2} - \alpha^2)  \\ 
\hat{V}(t) = \Omega_0 (t) \hat{a}_c \hat{a}_t^{\dagger} + h.c. \\
\end{array}
\label{eq:Control_Hamiltonian_ZZ}
\end{equation}
where $\Omega_0(t)$ is again real hard pulse.

We find that it is hard to apply derivative-based transition suppression technique with this original Hamiltonian since there are too many transition elements with distinct gap frequencies to be suppressed. Instead, it will be easier if we replace the BS interaction with two-mode squeezing $\hat{V}(t) = \Omega_0 (t) \hat{a}_c^{\dagger} \hat{a}_t^{\dagger} + h.c.$.

Again, working in Kerr-cat eigenbasis and only considering the first three pairs of eigenstates for each mode, we can express $\hat{H}_0$ and $\hat{V}$ as:
\begin{equation}
\begin{array}{c}
\hat{H}_0 = \Delta_1 (\hat{\Pi}_1^c + \hat{\Pi}_1^t) + \Delta_2 (\hat{\Pi}_2^c + \hat{\Pi}_2^t)\\
\hat{P}\hat{V}(t)\hat{P} = 2\alpha^2 \Omega_0 \hat{Z}_c\hat{Z}_t \otimes \hat{\Pi}_0 \\
 \hat{P}\hat{V}(t)\hat{Q} + \hat{Q}\hat{V}(t)\hat{P} = \Omega_0(t) \hat{Z}_c \hat{Z}_t \otimes [\alpha (\hat{\sigma}^{+}_{00,10} + \hat{\sigma}^{+}_{00,01}) + \eta \alpha (\hat{\sigma}^{+}_{00,20} + \hat{\sigma}^{+}_{00,02}) + \hat{\sigma}^{+}_{00,11} + \eta (\hat{\sigma}^+_{00,12} + \hat{\sigma}^+_{00,21}) + \eta^2\hat{\sigma}^+_{00,22}]  + h.c.
\end{array}
\end{equation}
where $\hat{\sigma}^+_{ij,i'j'} \equiv |i'\rangle_c |j'\rangle_t \langle i|_c \langle j|_t $. Here we only show the projection of $\hat{V}(t)$ in $\mathcal{H}_{logical}$ and its block off-diagonal part for simplicity.

\begin{table}[htp]
    \centering
    \begin{tabular}{p{3cm}<{\centering}|p{3cm}<{\centering}|p{3.5cm}<{\centering}|p{3cm}<{\centering}}
         \hline
         Transition & $\hat{Z}_c \hat{Z}_t \otimes( \hat{\sigma}^{+,c}_{0,1}, \hat{\sigma}^{+,t}_{0,1})$ & $\hat{Z}_c \hat{Z}_t \otimes (\hat{\sigma}^{+,t}_{0,2}, \hat{\sigma}^{+,c}_{0,2}, \hat{\sigma}^{+}_{00,11})$ & $\hat{Z}_c \hat{Z}_t \otimes(\hat{\sigma}^{+}_{00,12}, \hat{\sigma}^{+}_{00,21})$\\
         \hline
         Energy gap & $\Delta_a \equiv \Delta_1$ & $\Delta_b \equiv 2\Delta_1 (\Delta_2)$ & $\Delta_c \equiv \Delta_1 + \Delta_2$ \\
         \hline
    \end{tabular}
    \caption{The dominant diabatic transitions of the ZZ gate and their associated energy gaps. Here we use the fact that $\Delta_2 \approx 2\Delta_1$.}
    \label{tab:ZZ_gate_transitions}
\end{table}

Among all the diabatic transitions in the block off-diagonal part of $\hat{V}$ we consider suppressing three groups of them associated with three energy gaps $\Delta_a, \Delta_b, \Delta_c$ listed in Tab.~\ref{tab:ZZ_gate_transitions}. Similar to the Z rotation, to suppress the first-order leakage we first replace the base driving pulse as the Gaussian pulse with third-order smoothness $\Omega_0(t) = \Omega_{G,3}(t)$, then add $H_{DBC}$ by shaping the base driving pulse, i.e. $\hat{H}_{DBC} = u (t) \hat{a}_c \hat{a}_t + h.c.$. According to Eq.~\ref{eq:clssical_sol_pulse} the first-order correction pulse is:
\begin{equation}
	u^{(1)}(t) = -i \dot{\Omega}_{0}(t)\left(\frac{1}{\Delta_{a}}+\frac{1}{\Delta_{b}}+\frac{1}{\Delta_{c}}\right) -\ddot{\Omega}_{0}(t)\left(\frac{1}{\Delta_{a} \Delta_{b}}+\frac{1}{\Delta_{a} \Delta_{c}}+\frac{1}{\Delta_{b} \Delta_{c}}\right) + i\frac{\dddot{\Omega}_{0}(t)}{\Delta_{a} \Delta_{b} \Delta_{c}},
\end{equation}
and the corresponding first-order DRAG transformation is:
\begin{equation}
	\hat{S}^{(1)}(t) = i \Omega_0 \left(\frac{\hat{h}_a}{\Delta_{a}} +\frac{\hat{h}_b}{\Delta_{b}} + \frac{\hat{h}_c}{\Delta_{c}}\right) + \dot{\Omega}_0[\frac{1}{\Delta_a}(\frac{1}{\Delta_b} + \frac{1}{\Delta_c})\hat{h}_a + \frac{1}{\Delta_b}(\frac{1}{\Delta_a} + \frac{1}{\Delta_c})\hat{h}_b + \frac{1}{\Delta_c}(\frac{1}{\Delta_a} + \frac{1}{\Delta_b})\hat{h}_c] - \frac{\ddot{\Omega}_x}{\Delta_a \Delta_b \Delta_c}(\hat{h}_a + \hat{h}_b + \hat{h}_c) + h.c.
\end{equation}
where
\begin{equation}
\begin{array}{c}
\hat{h}_a = \hat{Z}_c \hat{Z}_t \otimes [\alpha (\hat{\sigma}^{+}_{00,10} + \hat{\sigma}^{+}_{00,10}) + \sqrt{2} \alpha (\hat{\sigma}^{+}_{01,02} + \hat{\sigma}^{+}_{10,20})] \\
\hat{h}_b = \eta \alpha \hat{Z}_c \hat{Z}_t \otimes [\hat{\sigma}^{+}_{00,11}  + \eta \alpha (\hat{\sigma}^{+}_{00,02} + \hat{\sigma}^{+}_{00,20}]\\
\hat{h}_c = \eta \hat{Z}_c \hat{Z}_t \otimes (\hat{\sigma}^+_{00,12} + \hat{\sigma}^+_{00,21}).
\end{array}
\end{equation}

Then the diabatic transitions are suppressed to the second order $O(|\mathcal{F}(\frac{\Omega_x^2}{\Delta}, \Delta, T)|^2)$. Similar to the Z rotation, there will be an over-rotation $\hat{Z}_c \hat{Z}_t \otimes \hat{\Pi}_0^c \hat{\Pi}_0^t$ coming from the higher-order DRAG expansion, which can be compensated by rescaling the rotation angle or adding an additional term to $u$:
\begin{equation}
u(t) = u^{(1)} + c_0 \frac{\Omega_0^3(t)}{\Delta_a^2},
\end{equation}
where we also numerically optimze $c_0$ and obtain $c_0 \approx 0.126$.

\subsection{CX gate}
The correction to the CX gate is more complicated and requires adding more physical correction terms to $H_{DBC}$ instead of merely pulse shaping.

The original Hamiltonian for the CX gate in \cite{Puri20} is:
\begin{equation}
\begin{array}{c}
\hat{H}_{original}(t) =\hat{H}^{(c)}_{\mathrm{KPO}}+\hat{H}^{(t)}_{\mathrm{KPO}}(t) + \hat{H}_{cp}\\
\hat{H}^{(c)}_{\mathrm{KPO}} =-K\left(\hat{a}_{\mathrm{c}}^{\dagger 2}-\alpha^{2}\right)\left(\hat{a}_{\mathrm{c}}^{2}-\alpha^{2}\right) \\
\hat{H}^{(t)}_{\mathrm{KPO}}(t) = -K\left[\hat{a}_{\mathrm{t}}^{\dagger 2}-\alpha^{2} e^{-2 i \phi(t)}\left(\frac{\alpha-\hat{a}_{\mathrm{c}}^{\dagger}}{2 \alpha}\right)-\alpha^{2}\left(\frac{\alpha+\hat{a}_{\mathrm{c}}^{\dagger}}{2 \alpha}\right)\right]\times\left[\hat{a}_{\mathrm{t}}^{2}-\alpha^{2} e^{2 i \phi(t)}\left(\frac{\alpha-\hat{a}_{\mathrm{c}}}{2 \alpha}\right)-\alpha^{2}\left(\frac{\alpha+\hat{a}_{\mathrm{c}}}{2 \alpha}\right)\right] \\
 H_{c p}=-\frac{1}{2} \dot{\phi} \frac{\left(2 \alpha-\hat{a}_{\mathrm{c}}^{\dagger}-\hat{a}_{\mathrm{c}}\right)}{2 \alpha} \left(\hat{a}_{\mathrm{t}}^{\dagger} \hat{a}_{\mathrm{t}}-\alpha^{2}\right).
\end{array}
\end{equation}

We first represent the operators on the control mode in the shifted fock basis and express $\hat{H}_{KPO}^{(t)}$ and $\hat{H}_{cp}$ as:
\begin{equation}
\begin{array}{c}
\hat{H}_{KPO}^{(t)} = - K [\hat{a}^{2\dagger}_t - \alpha^2 e^{- 2i\phi}\hat{P}^-_c - \alpha^2 \hat{P}^+_c + \frac{1}{2}\alpha (e^{-2i\phi} - 1)\hat{Z}_c \otimes \hat{a}_c^{'\dagger}] [\hat{a}^{2}_t - \alpha^2 e^{2i\phi}\hat{P}^-_c - \alpha^2 \hat{P}^+_c + \frac{1}{2}\alpha (e^{2i\phi} - 1)\hat{Z}_c \otimes \hat{a}_c'] \\
\hat{H}_{cp} = - \dot{\phi} [\hat{P}^-_c - \frac{1}{4\alpha} (\hat{a}'_c + \hat{a}'^{\dagger}_c)] (\hat{a}^{\dagger}_t \hat{a}_t - \alpha^2),
\end{array}{c}
\end{equation}
where $\hat{P}_c^+ \equiv \frac{\hat{I}_c + \hat{Z}_c}{2}$ and  $\hat{P}_c^- \equiv \frac{\hat{I}_c - \hat{Z}_c}{2}$.

We define the following adiabatic frame transformation 
\begin{equation}
\hat{U}(t)=\exp \left[-i \int_{0}^{t} dt' \dot{\phi}(t') \hat{P}_{c}^{-} \otimes(\hat{a}_{t}^{\dagger} \hat{a}_{t}-\alpha^{2})\right]=\hat{P}_{c}^{-} \otimes \exp \left[-i \phi(t)\left(\hat{a}_{t}^{\dagger} \hat{a}_{t}-\alpha^{2}\right)\right]+\hat{P}_{c}^{+},
\label{eq:Adiabatic_frame}
\end{equation}
and obtain the Hamiltonian in the adiabatic frame as:
\begin{equation}
\begin{array}{c}
\hat{\tilde{H}}_{original}(t)= \hat{U}(t)\hat{H}_{original}\hat{U}^{\dagger}(t) + i\dot{\hat{U}}(t)\hat{U}^{\dagger}(t) = \hat{\tilde{H}}_{0}+\hat{\tilde{H}}_{v}+\hat{\tilde{V}} \\
\hat{\tilde{H}}_{0}=-K \left(\hat{a}_{t}^{2 \dagger}-\alpha^{2}\right)\left(\hat{a}_{t}^{2}-\alpha^{2}\right)-K\left(\hat{a}_{c}^{2 \dagger}-\alpha^{2}\right)\left(\hat{a}_{c}^{2}-\alpha^{2}\right)\\
\hat{\tilde{H}}_{v}=  - \frac{1}{2} K \alpha\left[i \sin 2 \phi \hat{Z}_{c}-(1-\cos 2 \phi) \hat{I}_{c}\right] \otimes \hat{a}_{c}^{\prime} (\hat{a}_{t}^{2 \dagger}-\alpha^{2}) +h . c . - \frac{1}{2} K \alpha^{2}(1-\cos 2 \phi) \hat{I}_{c} \hat{a}_{c}^{\prime \dagger} \hat{a}_{c}^{\prime}\\
\hat{\tilde{V}}=\frac{1}{4 \alpha} \dot{\phi} \hat{Z}_{c}\otimes \left(\hat{a}_{c}^{\prime}+\hat{a}_{c}^{\prime \dagger}\right) \times(\hat{a}_{t}^{\dagger} \hat{a}_{t}-\alpha^{2})
\end{array}
\label{eq:H_CX_SF_basis}
\end{equation}

Only considering the first two pairs of eigenstates and expressing the annilation operator in the Kerr-cat eigenbasis as $\hat{a}_{c,t} = \hat{Z}_{c,t}\otimes(\alpha + \hat{\sigma}^{-,c,t}_{0,1} - \lambda_1 \hat{\Pi}_1^{c,t})$ in the Kerr-cat eigenbasis, we can write the Hamiltonian Eq.~\ref{eq:H_CX_SF_basis} as:

\begin{equation}
\begin{array}{c}
\hat{\tilde{H}}_0 = \Delta_1 \hat{I}_t \otimes \hat{\Pi}_1^{t}  + [\Delta_1 -  \frac{1}{2} K \alpha^{2}(1-\cos 2 \phi)] \hat{I}_c \otimes \hat{\Pi}_1^{c} + 2 K \alpha^2 \lambda_1^2 (1 - \cos 2\phi) \hat{I}_c \hat{I}_t \otimes \hat{\Pi}_1^c \hat{\Pi}_1^t\\
\hat{\tilde{H}}_{v} \approx  K \alpha^2 (1-\cos 2 \phi) \hat{I}_{c} \hat{I}_t \otimes  \left[ \hat{\sigma}^x_{01,10} - \lambda_1(\hat{\Pi}_{1}^{c} \hat{\sigma}_{0,1}^{x, t}+ \hat{\sigma}_{0,1}^{x, c} \hat{\Pi}_{1}^{t}) \right] -  K \alpha^2 \sin 2 \phi \hat{Z}_{c} \hat{I}_t \otimes \left[ \hat{\sigma}_{01,10}^y - \lambda_1 (\hat{\sigma}_{0,1}^{y, c}\hat{\Pi}_{1}^{t} - \hat{\Pi}_{1}^{c} \hat{\sigma}_{0,1}^{y, t}) \right]\\
\hat{\tilde{V}} \approx \frac{1}{4} \dot{\phi} \hat{Z}_{c} \hat{I}_{t}  \otimes \hat{\sigma}_{0.1}^{x, c} \hat{\sigma}_{0.1}^{x, t}.
\end{array}
\end{equation}

In this lab frame, the dominant leakage transition is $\hat{Z}_c \hat{I}_t \otimes \hat{\sigma}^+_{00,11}$ via $\hat{\tilde{V}}$ and its associated time-dependent energy gap is $\Delta_{11}(t) = 2 \Delta_1 - \frac{1}{2} K \alpha^2 (1 - 3\lambda_1^2)[1 - \cos 2\phi(t)]$. We can define the first-order DRAG transformation with respect to this transition:
\begin{equation}
\hat{S}^{(1)} = -\frac{1}{4} \frac{\dot{\phi}}{\Delta_{11}(t)}\hat{Z}_c\hat{I}_t \otimes \hat{\sigma}^y_{00,11},
\end{equation}
and obtain the first-order effective Hamiltonian in this DRAG frame:
\begin{equation}
\begin{aligned}
\hat{H}_{e f f}^{(1)} & =\hat{\tilde{V}}+i\left[\hat{S}^{(1)}, \hat{\tilde{H}}_{0}\right] + i\left[\hat{S}^{(1)}, \hat{\tilde{H}}_{v}\right] \\
& = \frac{1}{4} K \alpha (\alpha - \frac{1}{2}\lambda_1) \lambda_1 \frac{\dot{\phi} (1 - \cos 2\phi)}{\Delta_{11}(t)} \hat{Z}_c \hat{I}_t \otimes (\hat{\sigma}^{x,c}_{0,1}\hat{\Pi}_0^t + \hat{\Pi}_0^c \hat{\sigma}^{x,t}_{0,1}) - \frac{1}{4} K \alpha (\alpha - \frac{1}{2}\lambda_1) \lambda_1 \frac{\dot{\phi} \sin 2\phi}{\Delta_{11}(t)} \hat{I}_c \hat{I}_t \otimes (\hat{\Pi}_0^c \hat{\sigma}^{y,t}_{0,1} - \hat{\sigma}^{y,c}_{0,1}\hat{\Pi}_0^t) \\ 
& - \frac{1}{4} \frac{d}{dt}[\frac{\dot{\phi}}{\Delta_{11}(t)}] \hat{Z}_{c} \hat{I}_{t} \otimes \hat{\sigma}_{00,11}^{y}
\end{aligned}
\end{equation}.

To eliminate the $\hat{Z}_c \hat{I}_t \otimes \hat{\sigma}^+_{00,11}$ transition we can add the standard derivative correction $\hat{H}_{DBC, 0}^{(1)} = i u_{0}(t) \frac{\hat{a}_{c}-\hat{a}_{c}^{\dagger}}{4 \alpha} \times\left(\hat{a}_{t}^{\dagger} \hat{a}_{t}-\alpha^{2}\right)$, where $u_0(t) = \frac{d}{dt}[\frac{\dot{\phi}}{\Delta_{11}(t)}]$. 

Since $\hat{\tilde{H}}_{v}$ is not diagonal in the kerr-cat eigenbasis and is of order $\epsilon^0$, $i\left[\hat{S}^{(1)}, \hat{\tilde{H}}_{v}\right]$ adds new diabatic transitions to $H_{eff}^{(1)}$, which do not exist in the Lab frame. So we need to add additional terms to $H_{DBC}$ to suppress these transitions. 

To suppress $\hat{Z}_c \hat{I}_t \otimes \hat{\sigma}^{x,c}_{0,1}\hat{\Pi}_0^t$ we can add:
\begin{equation}
\hat{H}_{DBC,1}^{(1)} = c_1 \frac{\dot{\phi}(1-\cos 2 \phi)}{\Delta_{11}(t)} (\hat{a} + \hat{a}^{\dagger}),\ c_1 = \frac{1}{4} K \alpha (\alpha - \frac{1}{2}\lambda_1) \lambda_{1}.
\end{equation}

To suppress $\hat{I}_c \hat{I}_t \otimes \hat{\sigma}^{y,c}_{0,1}\hat{\Pi}_0^t$, we can add:
\begin{equation}
\hat{H}_{DBC,2}^{(1)} = i c_2 \frac{\dot{\phi} \sin 2 \phi}{\Delta_{11}(t)} \left(\hat{a}_{c}^{2}-\hat{a}_{c}^{2 \dagger}\right), \ c_2 = \frac{1}{8} K \alpha \lambda_{1}.
\end{equation}

And finally we can simultaneously suppress $\hat{Z}_c \hat{I}_t \otimes \hat{\Pi}_0^c \hat{\sigma}^{x,t}_{0,1}$ and $\hat{I}_c \hat{I}_t \otimes \hat{\Pi}_0^c \hat{\sigma} ^{y,t}_{0,1}$ by adding:
\begin{equation}
\hat{H}_{DBC,3}^{(1)} = c_3 \frac{\dot{\phi}(t)}{\Delta_{11}(t)} \left[\left(e^{2 i \phi(t)}-1\right) \hat{a}_{t}^{2 \dagger}+\left(e^{-2 i \phi(t)}-1\right) \hat{a}_{t}^{2}\right],\ c_3 =  \frac{1}{8} K \alpha \lambda_{1}.
\end{equation}

So in total we apply four derivative-based corrections $\hat{H}_{DBC}^{(1)} = \hat{H}_{DBC,0}^{(1)} + \hat{H}_{DBC,1}^{(1)} + \hat{H}_{DBC,2}^{(1)} + \hat{H}_{DBC,3}^{(1)}$ to suppress the first-order leakage to the first pair of excited states of each mode.

However, by adding $\hat{H}_{DBC,1}^{(1)}, \hat{H}_{DBC,3}^{(1)}$ we also induce extra unitary Z rotation on the control mode, which can be compensated by applying an additional Z rotation $\hat{Z} (\delta \theta)$ on control mode after the CX gate, which has negligible error compared to the CX gate. $\delta \theta$ can be calculated as:
\begin{equation}
\delta \theta = \frac{1}{4}K \alpha^2 \lambda_1 (3\alpha - \lambda_1) \int_{0}^T dt \frac{\dot{\phi}(t)[1-\cos 2 \phi(t)]}{\Delta_{11}(t)}.  
\end{equation}

\section{Details of the Numerical Simulations}
\begin{figure}[h!]
    \centering
    \includegraphics[width = 0.8\textwidth]{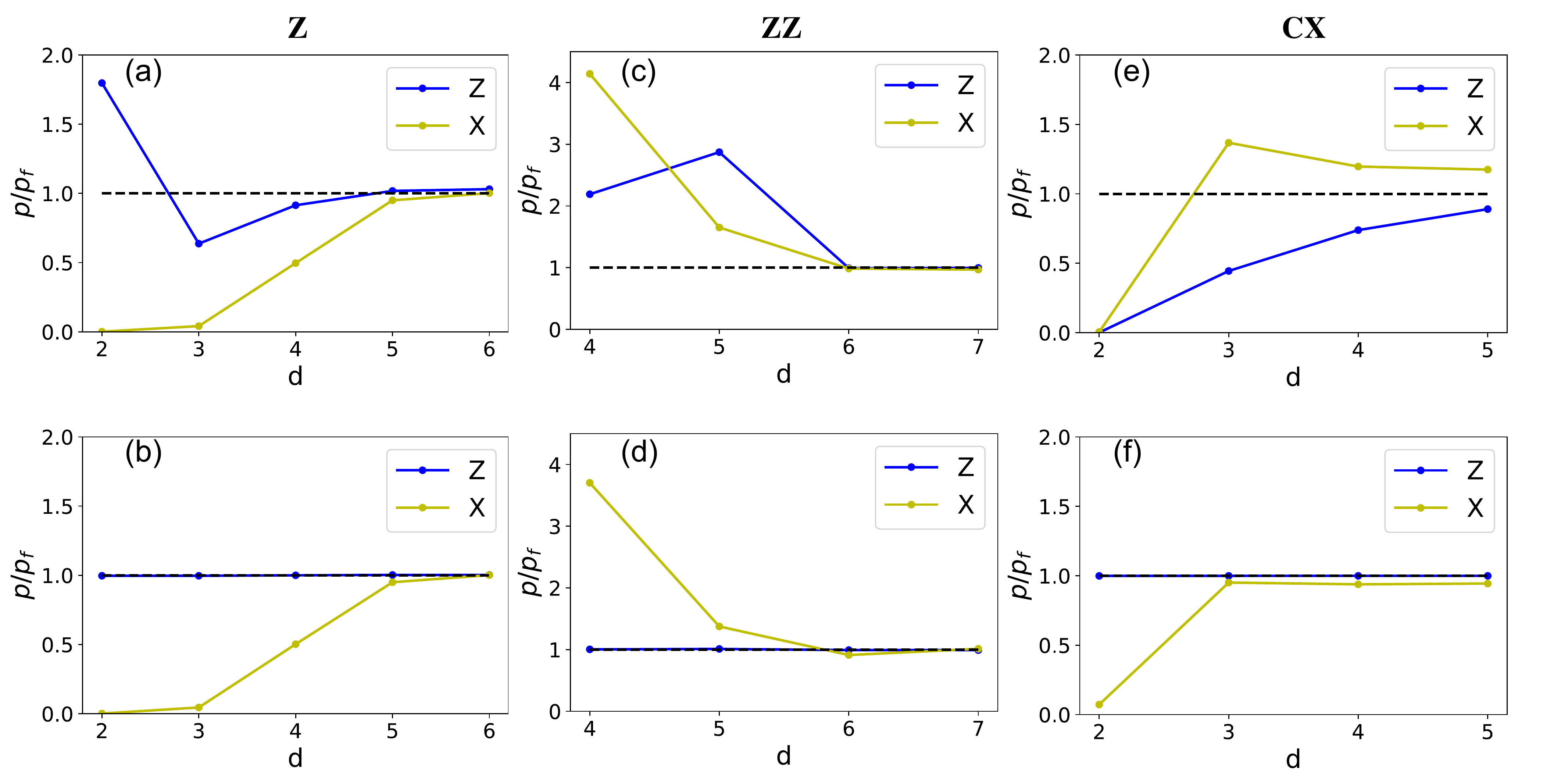}
    \caption{Convergency of the numerical simulation using Kerr-cat eigenbasis for $\alpha = \sqrt{8}$. (a)(c)(e) show the results when there is no photon loss. (b)(d)(f) show the results when there is photon loss with $\kappa_1/K = 10^{-4}$. The choice of the gate time for these three type of gates is: $T_Z = T_{ZZ} = 0.2/K$, $T_{cx} = 1/K$.}
    \label{fig:Convergency_Analysis}
\end{figure}

All simulations in this work use the Kerr-cat eigenbasis $\{|\psi_n^{\pm}\rangle \}$ with $n = 0,1,...,d - 1$, where $d$ is the truncation number and the dimension of the truncated Hilbert space for each mode is $2\times d$. The simulations of Z and ZZ rotation are in the lab frame. However, to simulate the CX gate using the Kerr-cat basis we need to move to the adiabatic frame defined in Eq.~\ref{eq:Adiabatic_frame}. The projector $\hat{P}^-_c$ is obtained approximately by:
\begin{equation}
\begin{aligned}
	\hat{P}^-_c & = \sum_{n=0}^{d - 1} \frac{1}{2} (|\psi_n^+ \rangle + (-1)^n |\psi_n^-\rangle) (\langle \psi_n^+ | + (-1)^n \langle \psi_n^-|) \\
	& = \sum_{n=0}^{d-1}|1\rangle_L \otimes |n''\rangle
\end{aligned}
\end{equation}
where the second line uses the subsystem decomposition.

To numerically extract the gate errors shown in the main text (Figs.2,3) we initialize the cats to certain initial states, simulate the gate dynamics, apply a strong two-photon dissipation ($\mathcal{D}[a^2 - \alpha^2]$) at the end of each gate and calculating the final state fidelity with target states. The application of the two-photon dissipation, which is a mathematical CPTP map that preserves the parity and locality in phase space, thus preserving the logical information, pushes all the leakage back into the cat state manifold so that we can explicitly evaluate the logical information. For simplicity, we choose $|+\rangle_L$/$|++\rangle_L$ as initial state to extract Z errors while $|0\rangle_L$/$|00\rangle_L$ to extract X errors. The choice of the above states for error extraction is justified in the next section where we provide the full error channel of the gates. 

The extracted gate errors using the Kerr-cat eigenbasis converge to those using the standard Fock basis as d increases. In Fig.~\ref{fig:Convergency_Analysis} we plot the ratio $\frac{P}{P_f}$ for different gates as a function of $d$, where $P$ is the total Z/X error probability obtained using the Kerr-cat eigenbasis and $P_f$ is the corresponding error probability obtained using the Fock basis. Based on Fig.~\ref{fig:Convergency_Analysis}, the value of d we choose to simulate Z rotation, ZZ rotation, CX gate are 5,6,4 respectively

\begin{figure}[h!]
    \centering
    \includegraphics[width = 0.9\textwidth]{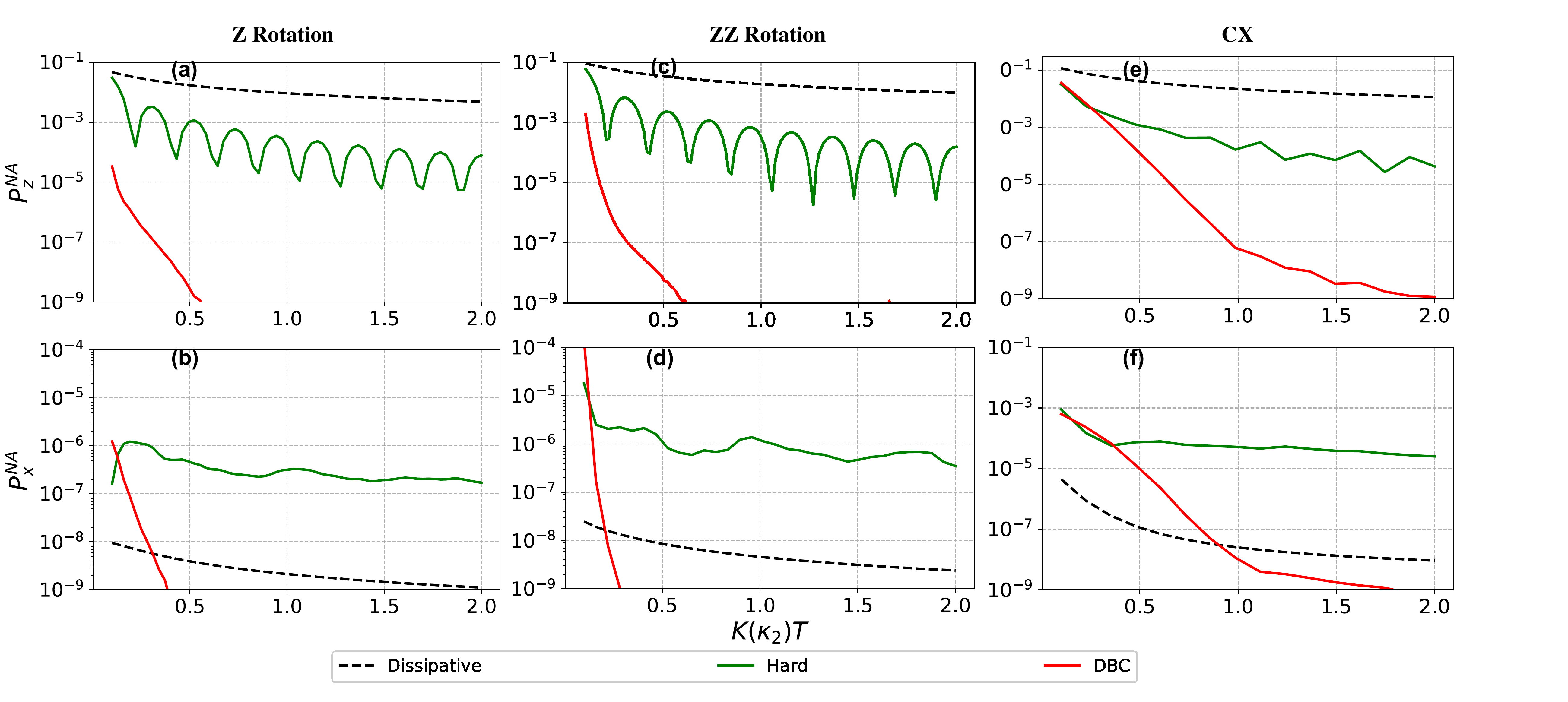}
    \caption{The non-adiabatic errors of all three gates considered in this work including the ZZ rotation which is not included in the main text.}
    \label{fig:non-adiabatic_error_all_gates}
\end{figure}

In Fig.~\ref{fig:non-adiabatic_error_all_gates} we show the numerically extracted non-adiabatic errors of all three type of gates considered in this work, including the ZZ rotation which is not covered in the main text. And in Tab.~\ref{tab:Noisy_error_scaling} we show the numerically fitted scalings of the minimal Z error probability of the CX gates with photon loss rate $\kappa_1$, which are in well agreement with the analysis in the main text.

\begin{table}[h]
    \centering
    \begin{tabular}{p{2.5cm}<{\centering}|p{1.5cm}<{\centering}|p{1.5cm}<{\centering}|p{1.5cm}<{\centering}}
         \hline
         Control Scheme & Dissipative & Hard & DBC \\
         \hline
         $P_z^* \propto (\frac{\kappa_1}{K(\kappa_2)})^p$  & $ p = 0.49$ & $ p = 0.68$ & $p = 0.84$\\
         \hline

    \end{tabular}
    \caption{The numerically fitted $\kappa_1$ dependence of the minimal Z error probability $P^*_z$ of CX gates with different control schemes.}
    \label{tab:Noisy_error_scaling}
\end{table}

\section{Error channel of the CX gate}
\begin{figure}[h!]
    \centering
    \includegraphics[width = 0.8\textwidth]{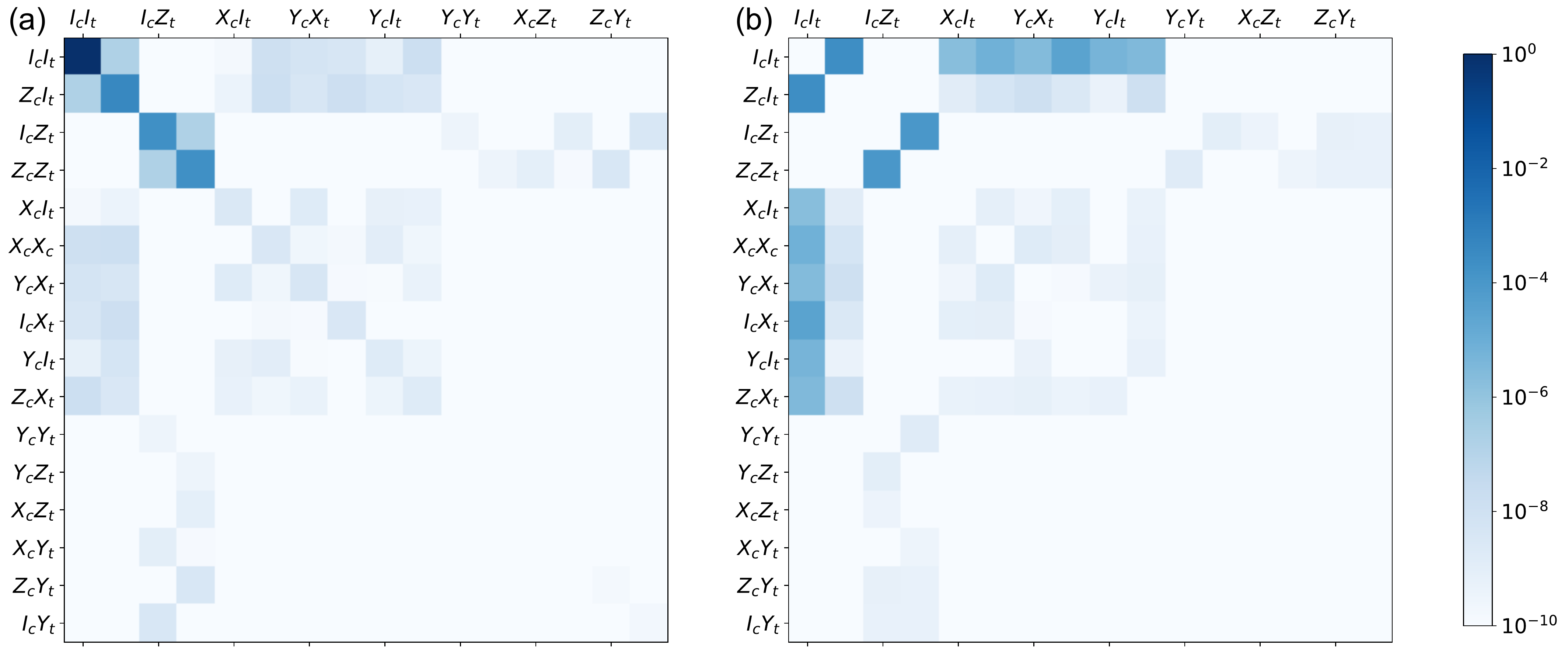}
    \caption{Numerically extracted error channel of the CX gate. The real part $|\rm{Re}(\chi^{err})|$ (a) and the imaginary part $|\rm{Im}(\chi^{err})|$ (b) of the error process matrix are plotted separately.}
    \label{fig:gate_tomo}
\end{figure}

Here we present the numerically extracted error channel of the Kerr CX gate in the presence of photon loss. In Fig.~\ref{fig:gate_tomo}(a)(b) we plot the real and imaginary part of the error matrix $\chi^{err}$ obtained from numerical process tomography, which describes the CPTP map:
\begin{equation}
    \mathcal{E}(\rho)=\sum_{m n} \chi_{m n}^{\mathrm{err}} P_{m} \tilde{\rho} P_{n}^{\dagger}
\end{equation}
where $\tilde{\rho} = CX \rho CX$ is the image of $\rho$ by an ideal CX gate, and $Pn$ ($n=0,1,2,3$) denotes the Pauli operators. The parameters used for the presented numeric result are: $\kappa_1/K = 5\times 10^{-5}, K T = 1$. The dominant errors are Z type of errors $Z_C, Z_t, Z_c Z_t$ while all non-Z type of errors are much smaller. We define the total Z error probability $p_Z^{tomo}$ as the sum of diagonal coefficients of $\chi^{err}$ corresponding to $Z_c, Z_t$ and $Z_c Z_t$ while the total X error probability $p_X^{tomo}$ as the sum of the rest of diagonal error coefficients. We compare $p_Z^{tomo}/p_X^{tomo}$ with the total error probability $p_Z/p_X$ extracted by calculating the target state infidelity with initial states $|++\rangle_L/|00\rangle_L$: $p_Z^{tomo} = 0.0008, p_Z = 0.0008; p_X^{tomo} = 1.71\times 10^{-8}, p_X = 1.41 \times 10^{-8}$. So the gate errors extracted using the chosen initial states well capture the error channel of the CX gate and the numerical results presented in this work are well justified.

\section{Details of the concatenated QEC}
\begin{figure}[h!]
    \centering
    \includegraphics[width = 0.45\textwidth]{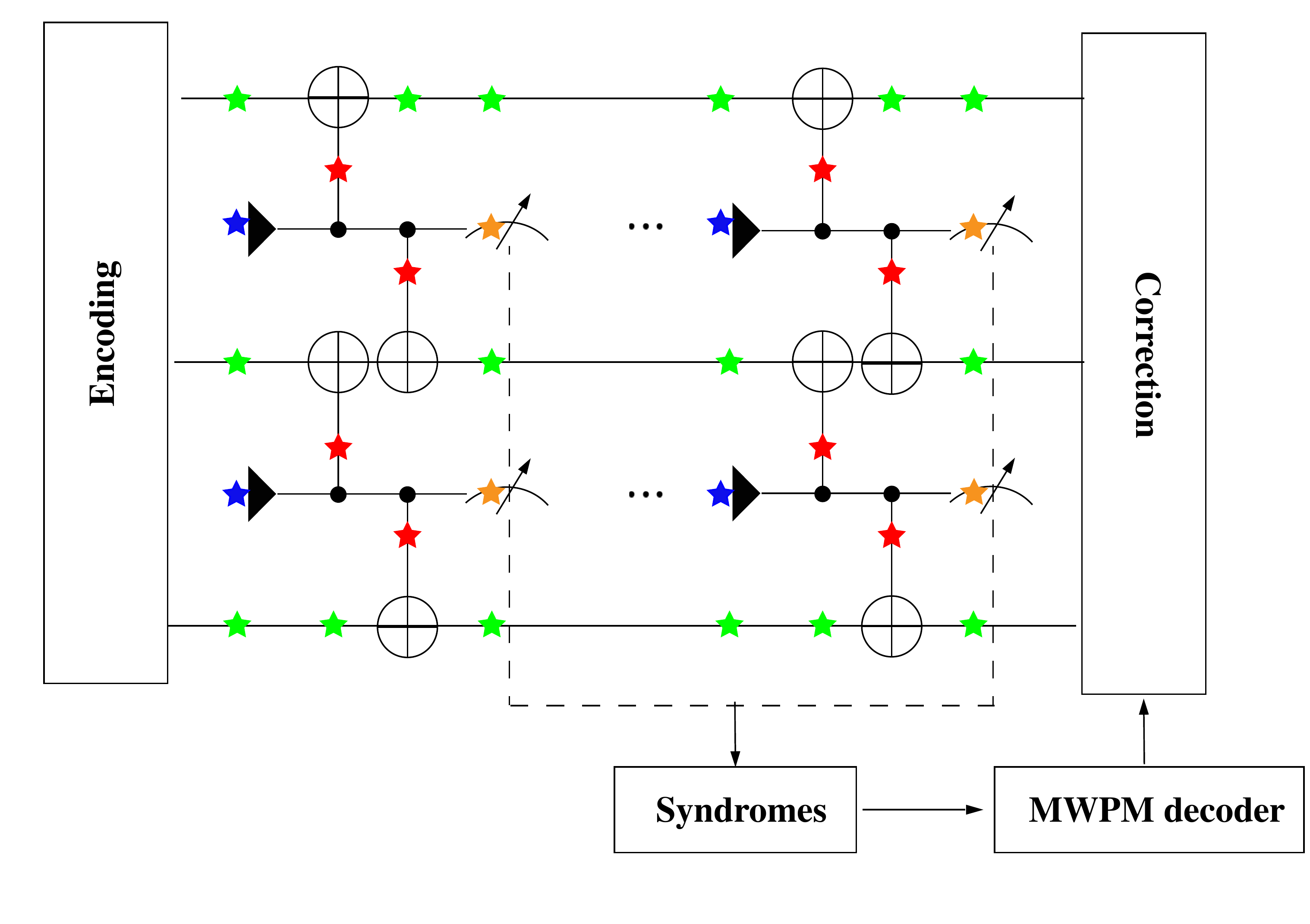}
    \caption{The QEC circuit of the repetition cat. The potential faulty operations are state preparation (red), idling (green), CX gate (red) and measurement (orange). The measured syndromes are feed into the minimum weight perfect matching (MWPM) decoder to determine the errors.}
    \label{fig:rep_circuit}
\end{figure}
% \section{Simulation of the repetition cat}
In this section we provide the details of the concatenated QEC considered in this work, including the detailed error model and how we obtain the logical Z error rate and the final optimal logical gate error rate.

The faulty operations we consider are state preparation, idling, measurement and CX gate marked by stars with different colors in the QEC circuit show in Fig.~\ref{fig:rep_circuit}. The physical Z and X error rates of these operations are summarized in Tab.~\ref{tab:error_model1}. The Z errors of the CX gate comprise of three parts: the Z error on the control mode which results from both non-adiabaticity and photon loss, the Z error on the target mode and correlated Z errors which are both induced by photon loss. $p_0 = \kappa_1 |\alpha|^2 T_{cx}$ is the characteristic photon-loss-induced Z error probability.

\begin{table}[h]
    \centering
    \begin{tabular}{p{1.5cm}<{\centering}|p{1cm}<{\centering}|p{1cm}<{\centering}|p{1cm}<{\centering}|p{3cm}<{\centering}}
         \hline
         Operation & Idle & $\mathcal{P}_{|+\rangle}$ & $\mathcal{M}_X$ & CX\\
         \hline 
         $P_z$ & $p_0$ & $p_0$ & $p_0$ & $Z_c: p^{NA}_{z}(T_{cx}) + p_0$ $Z_t: \frac{1}{2}p_0$; $Z_c Z_t: \frac{1}{2}p_0$\\
         \hline
         $P_x$ & 0 &  / & / & $p_x(\kappa_1, T_{cx})$ \\
         \hline
    \end{tabular}
    \caption{The physical error rates of different operations in our error model. $p_0 = \kappa_1 |\alpha|^2 T_{cx}$ is the characteristic Z error rate.}
    \label{tab:error_model1}
\end{table}

Using our model defined in Tab. ~\ref{tab:error_model1}, the logical Z error probability $P^L_Z$ of the logical CX gate will in general be a function of photon loss rate $\kappa_1$, repetition code distance $d$ and the physical CX gate time $T_{cx}$. In the following we will show how we estimate $P^L_Z$ in the low-loss regime while avoiding large Monte Carlo (MC) simulations.

We define a dimensionless parameter $\eta \equiv \frac{T_{cx}}{T^* }$ as the ratio between the chosen physical CX gate time $T_{cx}$ and the gate time $T^*$ that maximizes the physical CX gate fidelity. 

\begin{figure}[h!]
    \centering
    \includegraphics[width = 0.6\textwidth]{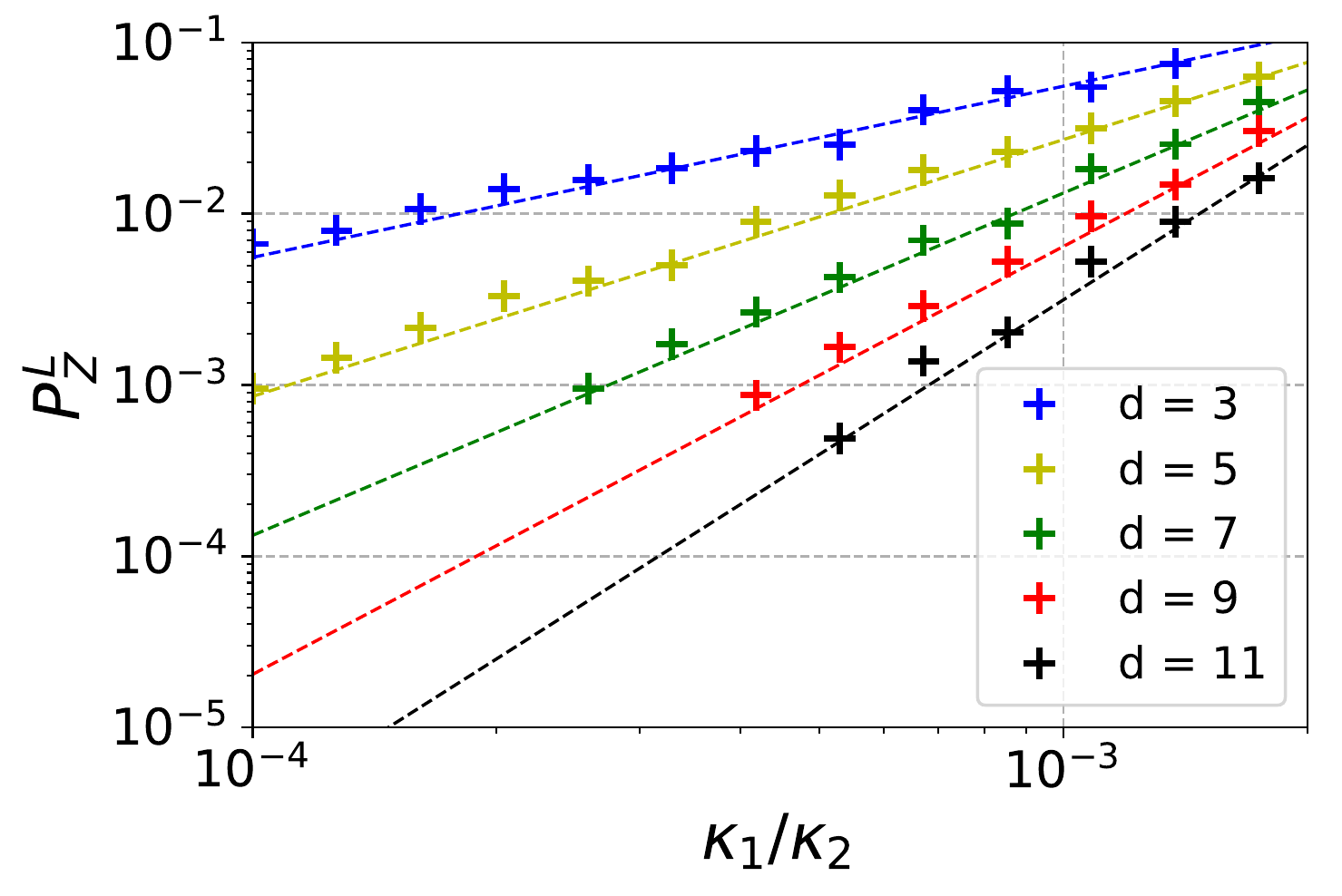}
    \caption{Logical Z error probability of the logical CX gate encoded in repetition-cat using dissipative CX gates. The '+' points are the MC simulation results while the dotted lines are the fitting results $P^L_Z = A [B \frac{\kappa_1}{\kappa_2}]^{\frac{d + 1}{4}}$. For this plot $\eta_T$ is set to 1 and the fitted coefficients are $A = 0.16$, $B = 352.1$. }
    \label{fig:2d_fitting}
\end{figure}

\begin{figure}[h!]
    \centering
    \includegraphics[width = 0.6\textwidth]{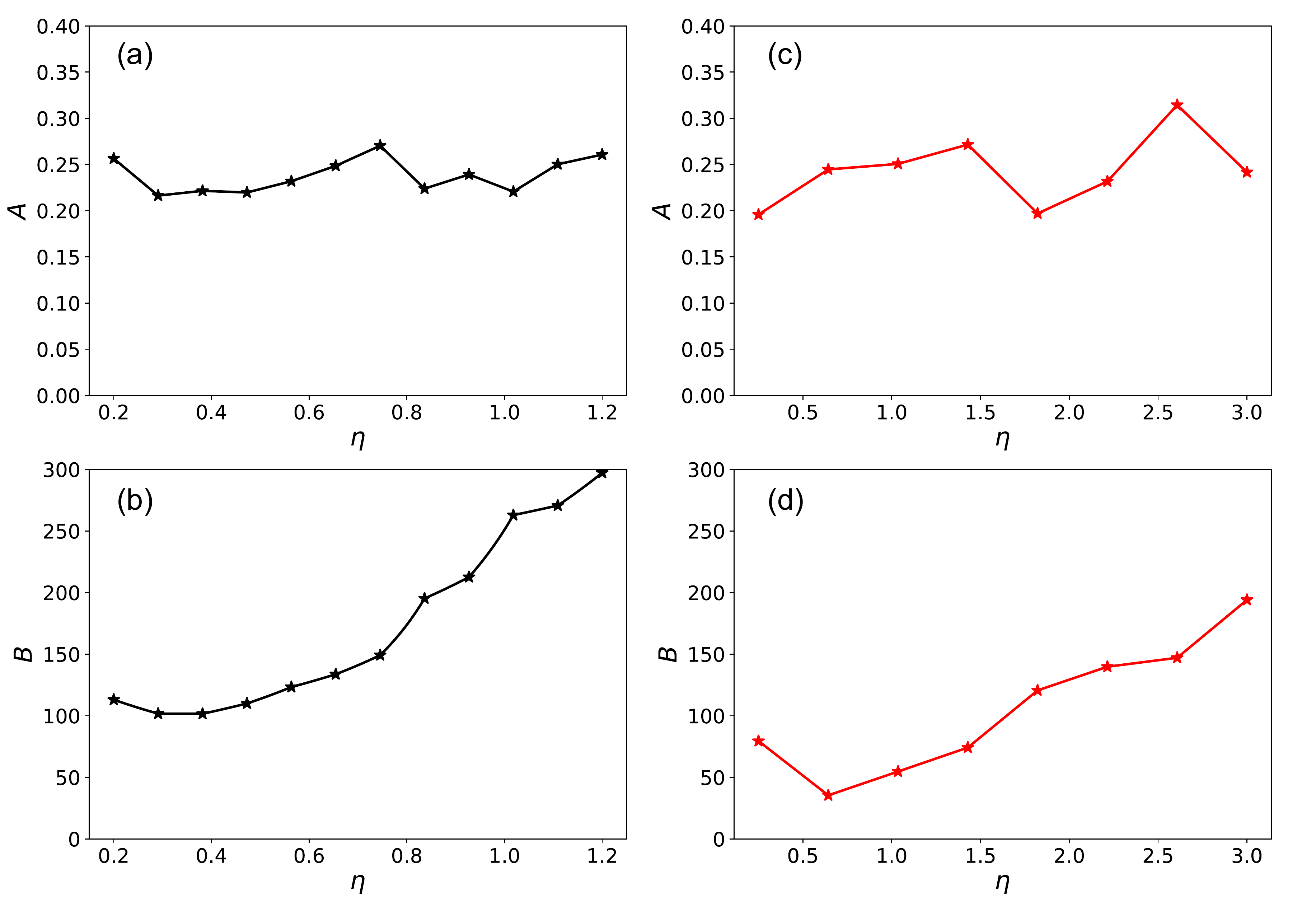}
    \caption{(a)(b): The fitted coefficients $A$ and $B$ in Eq.~\ref{eq:fitted_plz_dissi} as functions of $\eta$. (c)(d): The fitted coefficients $A$ and $B$ in Eq.~\ref{eq:fitted_plz_hard} as functions of $\eta$.}
    \label{fig:fitted_AB}
\end{figure}

 Using the dissipative CX gate, $T^* = \frac{\pi}{8\alpha^2 \sqrt{2\kappa_1 \kappa_2}}$ \cite{Chamberland_2020} and both the non-adiabatic error probability $P^{NA}_z$ and the characteristic Z error probability $p_0$ in Tab.III will be proportional to $\sqrt{\frac{\kappa_1}{\kappa_2}}$ for a given $\eta$. Therefore, we expect an empirical scaling of $P^L_Z$:
 \begin{equation}
 	P^L_Z(\kappa_1, \eta, d) = A (\eta) [B(\eta) \frac{\kappa_1}{\kappa_2}]^{\frac{d + 1}{4}}.
 	\label{eq:fitted_plz_dissi}
 \end{equation}
 This is numerically verified using the MC simulation and the coefficients $A(\eta)$ and $B(\eta)$ are fitted. As an example, the numerically obtained and fitted $P^L_Z$ for $\eta = 1$ is shown in Fig.~\ref{fig:2d_fitting}. The fitted coefficients $A(\eta)$ and $B(\eta)$ are shown in Fig.~\ref{fig:fitted_AB}(a)(b). By interpolating $A(\eta)$ and $B(\eta)$ using the numerically fitted data in Fig.~\ref{fig:fitted_AB}(a)(b) we can then use Eq.~\ref{eq:fitted_plz_dissi} to estimate the logical Z error probability $P^L_Z$ when choosing different dissipative CX gate time.   

Using physical CX gate on Kerr cat with hard control, $T^* = (\frac{\pi^2}{512 \alpha^6 \kappa_1 K^2})^{1/3}$ and both the non-adiabatic error probability $P^{NA}_z$ and the characteristic z error probability $p_0$ in Tab.III will be proportional to $(\frac{\kappa_1}{K})^{\frac{2}{3}}$ for a given $\eta$. So similarly we can fit the logical Z error probability as:

 \begin{equation}
 	P^L_Z(\kappa_1, \eta, d) = A (\eta) [B(\eta) \frac{\kappa_1}{K}]^{\frac{d + 1}{3}}.
 	\label{eq:fitted_plz_hard}
 \end{equation}
The fitted coefficients $A(\eta)$ and $B(\eta)$ are shown in Fig.~\ref{fig:fitted_AB}.(c)(d)

When using physical CX gate on Kerr cat with DBC control, we notice that for the range of gate time of interest, the non-adiabatic Z error probability $P_z^{NA}$ is negligible compared to the characteristic Z error probability $p_0$. So $P^L_Z$ is only a function of the $p_0$ and the code distance $d$, which can be numerically fitted as:
\begin{equation}
	P_{z}^{L}\left(\kappa_{1}, T_{c x}, d\right)=P_{z}^{L}\left(p_{0}\left(\kappa_{1}, T_{c x}\right), d\right) \approx 0.086 *\left(454 p_0\right)^{\frac{d+1}{2}},
	\label{eq:fitted_plz_DBC}
\end{equation}
where $p_0 = \kappa_1 \alpha^2 T_{cx}$.

\begin{figure}[h!]
    \centering
    \includegraphics[width = 0.7\textwidth]{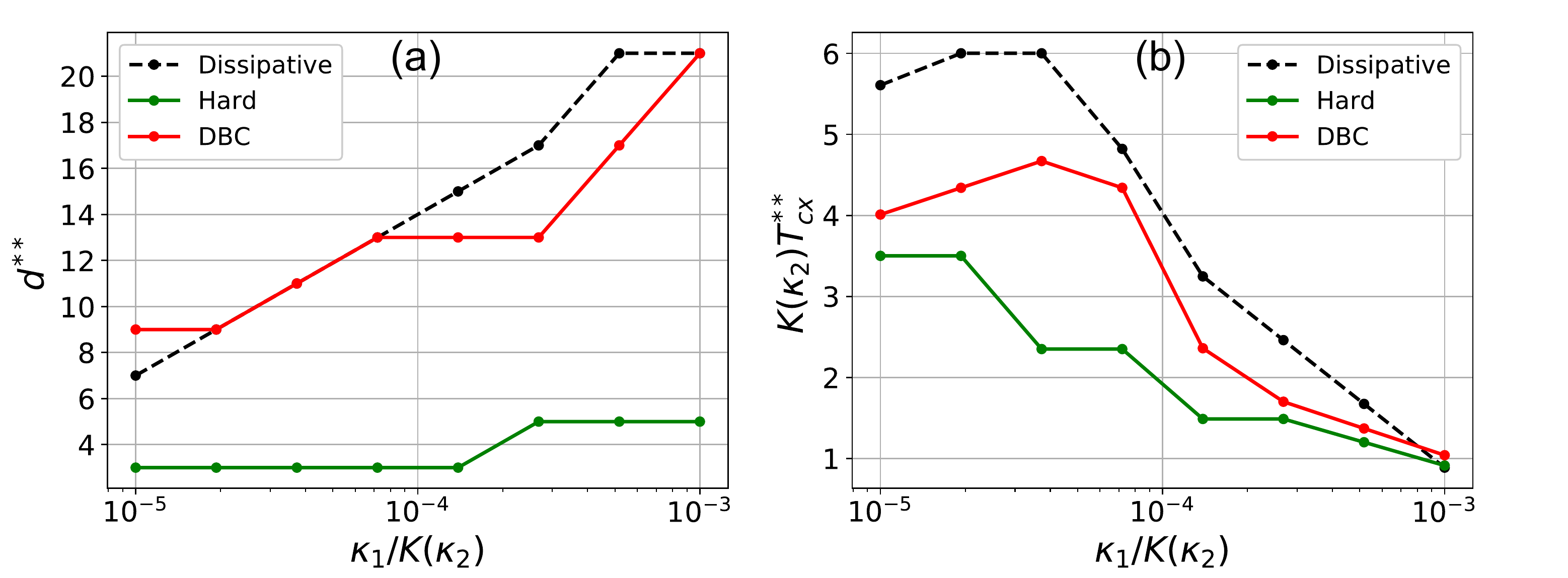}
    \caption{The optimal choice of code distance $d$ and physical CX gate time $T_{cx}$ to minimize the logical CX gate error.}
    \label{fig:optimal_d_T}
\end{figure}

In then end, we can obtain the minimal total logical error probability of the logical CX gate for given $\kappa_1$ by optimizing over $d$ and $T_{cx}$:
\begin{equation}
	P_{L}^{*}\left(\kappa_{1}\right)=\min _{T_{c x}, d} P^{L}\left(\kappa_{1}, T_{c x}, d\right) = \min _{T_{c x}, d} P^{L}_Z\left(\kappa_{1}, T_{c x}, d\right) + P^{L}_X\left(\kappa_{1}, T_{c x}, d\right),
\end{equation}
where $P^{L}_Z\left(\kappa_{1}, T_{c x}, d\right)$ is calculated using Eq.~\ref{eq:fitted_plz_dissi}~\ref{eq:fitted_plz_hard}~\ref{eq:fitted_plz_DBC}. The optimal choice of $d$ and $T_{cx}$ is shown in Fig.~\ref{fig:optimal_d_T}. We note that since the object function that we optimize is the total logical gate error rate, the optimal choice of repetition code distance $d^{**}$ here should not be considered as the resource overhead towards fault tolerance. In Fig.~\ref{fig:optimal_d_T}(a), the choice of $d^{**}$ using Hard control is small because the CX gates with Hard control have poor noise bias and consequently lead to high logical error rates (see Fig. 4 in the main text).

% % \input{SupplementaryMat.bbl}
% \bibliography{ref_supplementary} 
% \bibliographystyle{apsrev4-2}
% \bibliography{ref_supplementary} 
